\documentclass[final]{svjour2}
\usepackage{rotating}
\usepackage[colorlinks]{hyperref}
\hypersetup{colorlinks,citecolor=red,linkcolor=blue,urlcolor=blue}
\usepackage{mathptmx}
\usepackage[numbers, sort&compress]{natbib}
\usepackage{graphicx,amsmath,amssymb,color}
\graphicspath{{Figures/}}
\makeatletter \journalname{Journal of Low Temperature Physics}

\begin{document}

\newcommand{\hdblarrow}{H\makebox[0.9ex][l]{$\downdownarrows$}-}
\title{New state of matter: heavy-fermion systems, quantum spin
liquids, quasicrystals, cold gases, and high temperature
superconductors}

\author{V.R. Shaginyan\and V.A. Stephanovich\and A.Z. Msezane
\and P. Schuck \and J.W. Clark \and M. Ya. Amusia \and G.S.
Japaridze \and K.G.Popov\and E.V. Kirichenko}

\institute{V.R. Shaginyan \at Petersburg Nuclear Physics
Institute of NRC "Kurchatov Institute";
\\Gatchina, 188300, Russia, \at CTSPS, Clark Atlanta University,
\\ Atlanta, Georgia 30314, USA\\
\email{vrshag@thd.pnpi.spb.ru}
\\V.A. Stephanovich \at Institute of Physics, Opole University,
\\Oleska 48, 45-052, Opole, Poland\\ \email{stef@uni.opole.pl}
\\ A.Z. Msezane \at CTSPS, Clark Atlanta University,
\\ Atlanta, Georgia 30314, USA\\P.Schuck \at Institut de Physique Nucl\'eaire,
IN2P3-CNRS, Universit\'e Paris-Sud, F-91406 Orsay Cedex, France\\
J.W. Clark \at McDonnell Center for the Space Sciences \&
Department of Physics, Washington University, St.~Louis, MO
63130, USA; Center for Mathematical Sciences, University of
Madeira, Funchal, 9000-390 Portugal\\ M. Ya. Amusia \at Racah
Institute of Physics, Hebrew University, Jerusalem 91904,
Israel, \at Ioffe Physical Technical Institute, RAS, St.
Petersburg 194021, Russia\\ G. S.
Japaridze \at Clark Atlanta University, Atlanta, GA 30314, USA\\
K.G.Popov \at Komi Science Center, Ural Division, RAS, 3a,
Chernova str. Syktyvkar, 167982, Russia\\
E.V. Kirichenko \at Institute of Mathematics and
Informatics,Opole University, Oleska 48, 45-052, Opole, Poland}

\maketitle

\begin{abstract}

We report on a new state of matter manifested by strongly
correlated Fermi systems including various heavy-fermion (HF)
metals, two-dimensional quantum liquids such as  $\rm ^3He$
films, certain quasicrystals, and systems behaving as quantum
spin liquids. Generically, these systems can be viewed as HF
systems or HF compounds, in that they exhibit typical behavior
of HF metals.  At zero temperature, such systems can experience
a so-called fermion-condensation quantum phase transition
(FCQPT). Combining analytical considerations with arguments
based entirely on experimental grounds we argue and demonstrate
that the class of HF systems is characterized by universal
scaling behavior of their thermodynamic, transport, and
relaxation properties. That is, the quantum physics of different
HF compounds is found to be universal, emerging irrespective of
the individual details of their symmetries, interactions, and
microscopic structure. This observed universal behavior reveals
the existence of a new state of matter manifest in HF compounds.
We propose a simple, realistic model to study the appearance of
flat bands in two-dimensional ensembles of ultracold fermionic
atoms, interacting with coherent resonant light. It is shown
that signatures of these flat bands may be found in
peculiarities in their thermodynamic and spectroscopic
properties. We also show that the FCQPT, in generating flat
bands and altering Fermi surface topology, is an essential
progenitor of the exotic behavior of the overdoped
high-temperature superconductors represented by $\rm
La_{2-x}Sr_xCuO_4$, whose superconductivity differs from that
predicted by the classical Bardeen-Cooper-Schrieffer theory. The
theoretical results presented are in good agreement with recent
experimental observations, closing the colossal gap between
these empirical findings and Bardeen-Cooper-Schrieffer-like
theories.

\end{abstract}

\section{Introduction} \label{INTR}

Strongly correlated Fermi systems, such as heavy fermion (HF)
metals, high-$T_c$ superconductors, two-dimensional liquids like
$\rm ^3He$, compounds with quantum spin liquids, quasicrystals,
and systems with one-dimensional quantum spin liquid, are among
the most intriguing and thoroughly experimentally studied
fundamental systems in physics. The properties of these
materials, called HF compounds, differ dramatically from
those of ordinary systems of interacting fermions, well described within the framework
of famous Landau Fermi liquid (LFL) theory, see e.g.
\cite{ste,varma,vojta,voj,belkop,obz,cust,sen,senth1,senth2,senth3,col11,col2,geg1,geg,huy,steg,col1}.
For instance, in the case of metals with HF, the strong
electron-electron correlations lead to renormalization of the
quasiparticles effective mass $M^*$, which may exceed the
ordinary, "bare", mass by several orders of magnitude or even
become infinitely large. The effective mass strongly depends on
the temperature, pressure, or applied magnetic field. Such
metals exhibit NFL behavior and unusual power laws of the
temperature dependence of the thermodynamic properties at low
temperatures.

The Landau theory of the Fermi liquid is based on the mapping of the
system (liquid) of strongly interacting electrons and nuclei to that of a
weakly interacting Fermi gas. This implies that the elementary excitations
behave as quasiparticles of a weakly interacting Fermi gas, determining the
system physical properties at low temperatures.
Thus latter excitations have a certain effective mass $M^*$, which depends
weakly on temperature,
pressure, and magnetic field strength, and is a parameter of the
theory \cite{landau,lanl1,PinNoz}. The LFL theory fails to
explain the results of experimental observations related to the
dependence of $M^*$ on the temperature $T$, magnetic field $B$,
pressure and other external stimuli. This led to the evasive conclusion that
quasiparticles do not survive in strongly correlated Fermi
systems and that the heavy electron does not retain its identity
as a quasiparticle excitation
\cite{cust,sen,senth1,senth2,senth3,col11,col2,col1}.

The unusual properties and NFL behavior observed in HF compounds
are assumed to be determined by various magnetic quantum phase
transitions
\cite{ste,varma,vojta,voj,belkop,obz,cust,sen,senth1,senth3,col11,col2,geg1}.
Since a quantum phase transition occurs at the temperature
$T=0$, the control parameters are the composition, electron
(hole) number density $x$, pressure, magnetic field strength
$B$, etc. A quantum phase transition occurs at a quantum
critical point, which separates the ordered phase that emerges
as a result of quantum phase transition from the disordered
phase. It is usually assumed that magnetic (e.g., ferromagnetic
and antiferromagnetic) quantum phase transitions are responsible
for the NFL behavior. The critical point of such a phase
transition can be shifted to absolute zero by varying the above
parameters. The observed universal behavior can be expected only
if the system under consideration is very close to a quantum
critical point, e.g., when the correlation length is much longer
than the microscopic length scale, and critical quantum and
thermal fluctuations determine the anomalous contribution to the
thermodynamic functions of the metal. Quantum phase transitions
of this type are so widespread
\cite{varma,vojta,voj,senth1,senth2,senth3,col11,col2} that we
call them ordinary quantum phase transitions
\cite{shag3,pr,book}. In this case, the physics of the
phenomenon is determined by thermal and quantum fluctuations of
the critical state, while quasiparticle excitations are
destroyed by these fluctuations. Conventional arguments that
quasiparticles in strongly correlated Fermi liquids "get heavy
and die" at a quantum critical point commonly employ the
well-known formula based on the assumptions that the $z$-factor
(the quasiparticle weight in the single-particle state) vanishes
at the points of second-order phase transitions \cite{col1}.
However, it has been shown that this scenario is problematic
\cite{pr,book,khodb,x18}, while in the case of the heavy-fermion
(HF) metal $\rm \beta-YbAlB_4$ experimental facts show that
magnetic quantum phase transitions and the corresponding
fluctuations are not responsible for the observed NFL behavior
\cite{pr,book,s2015,ybalb}. The same conclusion is true in the
case of compounds with strongly correlated quantum spin liquids,
two-dimensional liquids like $\rm ^3He$, quasicrystals, and
systems with one-dimensional quantum spin liquid
\cite{prl3he,shaginyan:2011,shaginyan:2012:A,shaginyan:2011:C,quasicryst,sc2016,annphys}.

Another difficulty is in explaining the restoration of the LFL
behavior under the application of magnetic field $B$, as
observed in HF metals and in high-$T_c$ superconductors
\cite{ste,geg,cyr} and the other HF compounds, see Section
\ref{QSL}. For the LFL state as $T\to0$, the electric
resistivity $\rho(T)=\rho_0+AT^2$, the heat capacity
$C(T)=\gamma_0T$, and the magnetic susceptibility $\chi=const$.
It turns out that the coefficient $A(B)$, the Sommerfeld
coefficient $\gamma_0(B)\propto M^*$, and the magnetic
susceptibility $\chi(B)$ depend on the magnetic field strength B
such that $A(B)\propto\gamma_0^2(B)$ and $A(B)\propto\chi^2(B)$,
which implies that the Kadowaki-Woods relation
$K=A(B)/\gamma_0^2(B)$ \cite{kadw} is $B$-independent and is
preserved \cite{geg}. Such universal behavior, quite natural
when quasiparticles with the effective mass $M^*$ playing the
main role, can hardly be explained within the framework of the
approach that presupposes the absence of quasiparticles, which
is characteristic of ordinary quantum phase transitions in the
vicinity of QCP. Indeed, there is no reason to expect that
$\gamma_0$, $\chi$ and $A$ are affected by the fluctuations in a
correlated fashion. For instance, the Kadowaki-Woods relation
does not agree with the spin density wave scenario \cite{geg}
and with the results of research in quantum criticality based on
the renormalization-group approach \cite{mill}. Moreover,
measurements of charge and heat transfer have shown that the
Wiedemann-Franz law holds in some high-$T_c$ superconductors
\cite{cyr,cyr1} and HF metals \cite{pag,pag2,ronn1,ronn}. All
this suggests that quasiparticles do exist in HF compounds, and
this conclusion is also corroborated by photoemission
spectroscopy results and measurements of the Kadowaki-Woods
relation in such uncommon HF compound as quasicrystals, see e.g.
\cite{koral,fujim,QCM}. We show that the basic properties of HF
compounds can be described within the framework of a fermion
condensation quantum phase transition (FCQPT), leading to
formation of fermion condensation (FC) and flat bands,  and
extended quasiparticle paradigm that allow us to explain the
non-Fermi liquid behavior observed in strongly correlated Fermi
systems \cite{pr,book}. In contrast to the Landau paradigm
assuming that the quasiparticle effective mass is approximately
constant, the effective mass of new quasiparticles strongly
depends on temperature, magnetic field, pressure, and other
parameters.  {We note that the direct experimental manifestation
of FC has been done recently \cite{mel2016}.}

The rest of the paper is organized as follows: In Section
\ref{scal}, we use HF metals as an example, and outline the
scaling behavior observed in HF compounds. Upon introducing
internal scales, we show that the behavior is universal for HF
compounds. In Sections \ref{QSL} we examine the universal
scaling behavior of the thermodynamic, transport and relaxation
properties of HF compounds magnets of new types with strongly
correlated quantum spin liquid (SCQSL), the recently discovered
quasicrystals, respectively, and show that these HF compounds
demonstrate the new state of matter. Section \ref{gas} presents
a perspective simple realistic model to observe the appearance
of flat bands in two-dimensional ensemble of ultracold fermionic
atoms, interacting with coherent resonant light. In Section
\ref{hts}, we show that FCQPT, generating flat bands and
altering Fermi surface topology, is a primary reason for the
recently observed behavior of the overdoped high-temperature
superconductors represented by $\rm La_{2-x}Sr_xCuO_4$, whose
superconductivity features differ from what is predicted by the
classical Bardeen-Cooper-Schrieffer theory. Section \ref{SUM}
summaries the main results, stressing the observation that the
quantum physics of different HF compounds is universal and
emerges regardless of their underlying microscopic details. This
uniform behavior, formed by flat bands, manifests the new state
of matter.

\section{Scaling behavior and internal scales}\label{scal}

\begin{figure} [! ht]
\begin{center}
\vspace*{-0.2cm}
\includegraphics [width=1.0\textwidth]{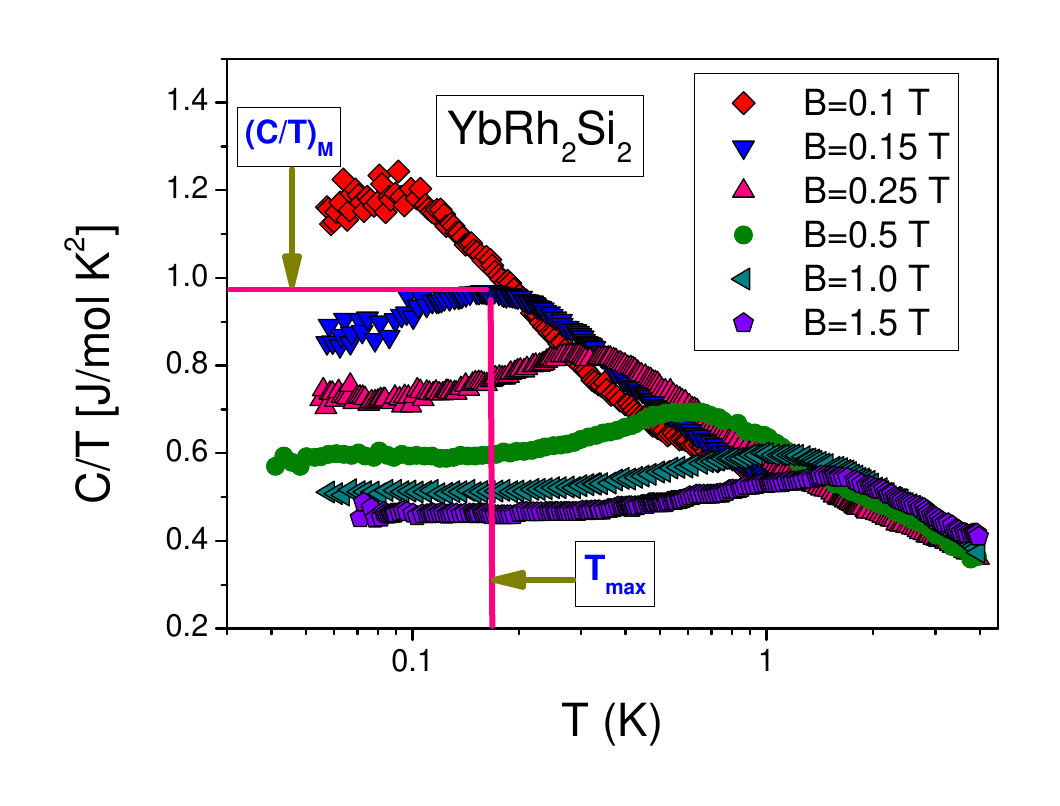}
\end{center}
\vspace*{-0.3cm} \caption{(Color online) Electronic specific
heat of $\rm YbRh_2Si_2$, $C/T$, versus temperature $T$ as a
function of magnetic field $B$ \cite{oesb} shown in the
legend.}\label{YBRHSI}
\end{figure}

As we have mentioned in Section \ref{INTR}, the NFL behavior
manifests itself in the power-law behavior of the physical
quantities of HF compounds, with exponents different from those
of a Fermi liquid \cite{oesb,oesbs}. It is common belief that
the main output of theory is the explanation of these exponents
which are at least depended on the magnetic character of QCP and
dimensionality of the system. On the other hand, the NFL
behavior cannot be captured by these exponents as seen from Fig.
\ref{YBRHSI}. Indeed, as a function of $T$ at fixed $B$, the
specific heat $C/T$ exhibits a behavior that is to be described
as a function of both temperature $T$ and magnetic $B$ field
rather than by a single exponent. One can see that at low
temperatures $C/T$ demonstrates the LFL behavior which is
changed by the transition regime at which $C/T$ reaches its
maximum and finally $C/T$ decays into NFL behavior. It is seen
from Fig. \ref{YBRHSI} that, both being check in the LFL regime
and in the transition one, these exponents may have little
physical significance.

\begin{figure} [! ht]
\begin{center}
\vspace*{-0.2cm}
\includegraphics [width=1.0\textwidth]{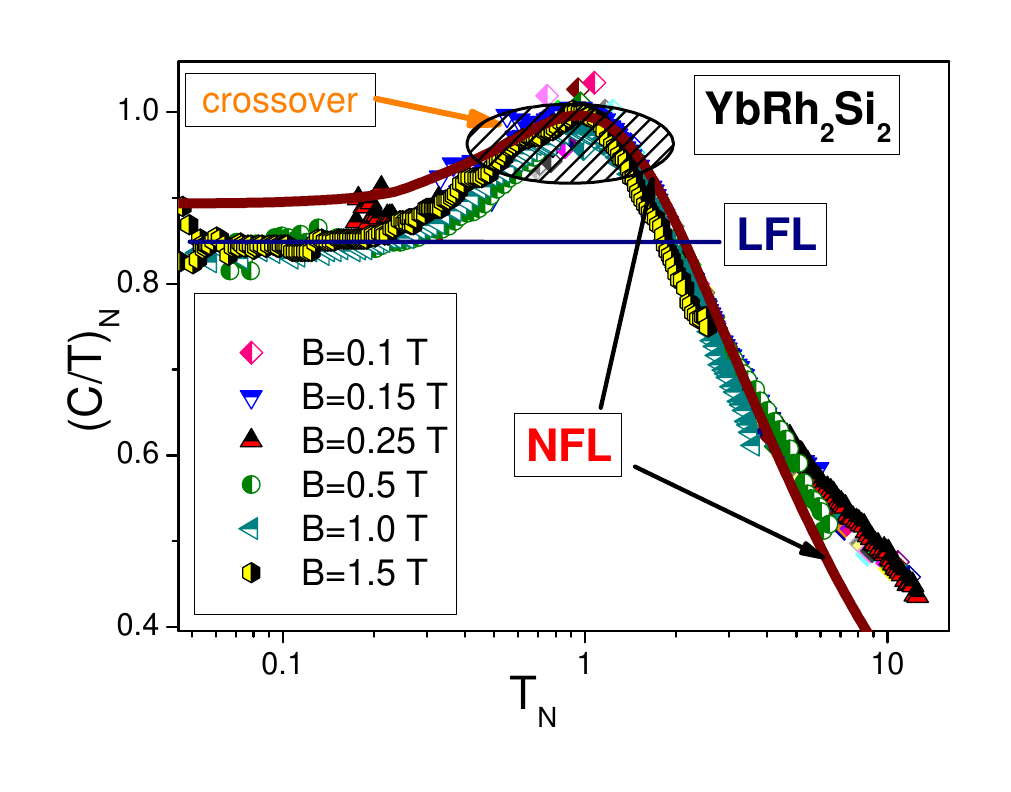}
\end{center}
\vspace*{-0.3cm} \caption{(Color online) The normalized
effective mass $M^*_N$ versus normalized temperature $T_N$.
$M^*_N$ is extracted from the measurements of the specific heat
$C/T$ on $\rm YbRh_2Si_2$ in magnetic fields $B$ \cite{oesb}
listed in the legend. Constant effective mass inherent in normal
Landau Fermi liquids is depicted by the solid line. The NFL
regime is shown by the arrows. The crossover is depicted by the
shaded area. The solid curve represents our calculation of the
universal behavior of $(C/T)_N$.}\label{YBRHSIN}
\end{figure}

In order to reveal the universal scaling behavior, and to
establish that salient properties of $C/T$ displayed in
Fig.~\ref{YBRHSI} possess generic character, we recall that can
be helpful to use ``internal'' scales to measure the effective
mass $M^*\propto C/T$ and temperature $T$
\cite{pr,book,dft373,dftjtp}.  As successively higher magnetic
fields are applied to the sample, Fig.~\ref{YBRHSI} shows
successive maxima $(C/T)_M$ in $C/T$ at corresponding
temperatures $T_M$, with $T_M$ shifting to higher $T$ as $B$ is
increased.  The value of the Sommerfeld coefficient
$C/T=\gamma_0$ is saturated towards lower temperatures,
decreasing at elevated magnetic fields.  To reveal a remarkable
universal scaling behavior that is present independently of
``incidental'' system properties, we may adopt $(C/T)_M$ and
$T_M$ as relevant internal scales. Accordingly, the maximum
structure $(C/T)_M$ in $C/T$ is used to normalize $C/T$, while
$T$ is normalized by $T_M$.  The resulting plots of the
normalized $(C/T)_N =(C/T)/(C/T)_M$ versus the normalized
temperature variable $y=T_N=T/T_M$ fields are displayed in
Fig.~\ref{YBRHSIN}. The normalized experimental results for the
different magnetic fields $B$ are seen to merge into a single
curve over an extended range in $y$, providing clear
documentation of universal scaling behavior, with LFL and NFL
regimes separated by a transition region in which $(C/T)_N$
attains its maximum value.

\begin{figure} [! ht]
\begin{center}
\includegraphics [width=0.88\textwidth]{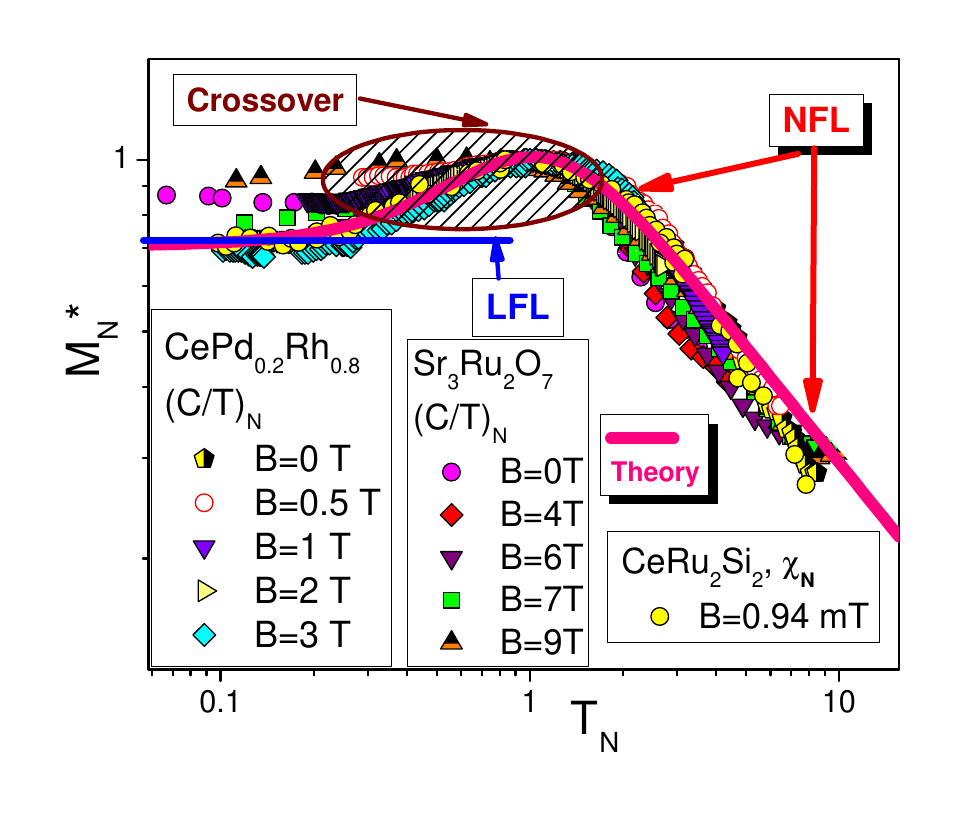}
\end{center}
\caption{The universal scaling behavior of the normalized
effective mass $M^*_N$ versus $T_N$, extracted from the
measurements of $\chi$ and $C/T$, see Eq. \eqref{NORM}, (in
magnetic fields $B$ shown in the legends) on $\rm CeRu_2Si_2$
\cite{takahashi:2003}, $\rm CePd_{1-x}Rh_x$ with $x=0.80$
\cite{oesbs}, and $\rm Sr_3Ru_2O_7$ \cite{rost:2011}. The LFL
and NFL regimes are shown by the arrows. The crossover is
depicted by the shaded area. The solid curve represents our
calculation of the universal behavior of $M^*_N(T_N)$, see Eq.
\eqref{UN2}.}\label{fig003}
\end{figure}

To analyze dependence of the effective mass $M^*$ on temperature
$T$, magnetic field $B$, momentum ${ p}$, number density $x$
etc., we use the Landau equation for the effective mass $M^*$
 \cite{landau,lanl1,PinNoz}
\begin{equation}\label{FLL} \frac{1}{M^*(T,B)} =
\frac{1}{M}+\int \frac{{\bf p}_F{\bf p}_1}{p_F^3} F({\bf p}_F,{
\bf p}_1,n)\frac{\partial n({ p_1},T,B)}{\partial { p_1}}
\frac{d{\bf p}_1}{(2\pi)^3},
\end{equation}
expressed in terms of the bare mass $M$, the Landau interaction
$F$, and the Fermi-Dirac distribution
\begin{equation} n_{\pm}({
p},T,B)= \left\{1+\exp\left[\frac{\varepsilon({ p},T)\pm
B\mu_B-\mu} {T}\right]\right\}^{-1}.\label{FL4}
\end{equation}
Here, $n_{\pm}({ p},T)$ and $\varepsilon({ p},T)$ are
respectively the the quasiparticle momentum distribution and
$\varepsilon$ the quasiparticle energy, with $\mu$ the chemical
potential and $\mu_B$ the Bohr magneton.  The term $\pm B\mu_B$
entering the right side of Eq.~\eqref{FL4} describes the Zeeman
splitting. Equation \eqref{FLL} is exact and can be derived
within the framework of the Density Functional Theory
\cite{pr,book}. This equation allows us to calculate the
behavior of $M^*$ which now becomes a function of temperature
$T$, external magnetic field $B$, number density $x$, pressure
$P$, etc. Near FCQPT the normalized solution of Eq. \eqref{FLL}
$M^*_N(T_N)$ can be approximated well by a simple universal
interpolating function \cite{pr,book}. The interpolation occurs
between the LFL and NFL regimes and represents the universal
scaling behavior of $M^*_N$
\begin{equation}M^*_N(T_N)\approx c_0\frac{1+c_1T_N^2}{1+c_2T_N^{n}}.
\label{UN2}
\end{equation}
Here, $c_0=(1+c_2)/(1+c_1)$, $c_1$, $c_2$ are fitting
parameters; and the exponent $n=8/3$ if the Landau interaction
is an analytical function, otherwise $n=5/2$. It follows from
Eq.~\eqref{FLL} that
\begin{equation}
\label{TMB} T_M\simeq a_1\mu_BB,
\end{equation}
where $a_1$ is a dimensionless number.  The result \eqref{TMB}
is in good agreement with experiment, although possible
corrections to it near the corresponding phase transitions are
discussed in Refs.~\cite{pr,book}. Importantly, the effective
mass $M^*$ defines the thermodynamic properties of HF compounds;
schematically, $M^*(T)\propto C(T)/T\propto S(T)/T\propto
\chi(T)$, where $C(T)$ is the specific heat, $S(T)$ the entropy,
$M_0(T)$ the magnetization, and $\chi(T)$ the AC magnetic
susceptibility. For the normalized values one has
\begin{equation}
\label{NORM} M^*_N=(C/T)_N=(S/T)_N=\chi_N.
\end{equation}

As seen from Fig. \ref{YBRHSIN}, the normalized effective mass
$(C/T)_N=M^*_N(T_N)$ extracted from the measurements is not a
constant, as would be for a LFL, and shows the scaling behavior
over three decades in normalized temperature $T_N$. It is seen
from Figs. \ref{YBRHSI} and \ref{YBRHSIN} that the NFL behavior
and the associated scaling extend at least to temperatures up to
few Kelvins. Scenario where fluctuations in the order parameter
of an infinite (or sufficiently large) correlation length and an
infinite correlation time (or sufficiently large) develop the
NFL behavior can hardly match up such high temperatures, while
the existence of quasiparticles completely account for this
behavior \cite{pr,book,qp1,qp2}.

We are led to conclude that a central challenge for mechanistic
theories of the critical behavior of the HF compounds, including
HF metals, lies in explanation of the scaling behavior of
$M^*_N(T_N)$. Theories calculating only the exponents
characterizing  $M^*_N(T_N)$ at $T_N\gg 1$ deal only with a part
of the observations and overlook, for example, what is happening
in the transition regime. Another qualitative aspect the problem
calling for explanation is the remarkably large temperature
ranges over which the NFL behavior is found to occur.
Fig.~\ref{fig003} informs us that this key feature may in fact
be understood within the theoretical framework of the fermion
condensate (FC).

The behavior of the normalized effective mass extracted from
measurements of $\chi$ and $(C/T)$ in the compounds $\rm
CeRu_2Si_2$ \cite{takahashi:2003}, $\rm CePd_{0.8}Rh_{0.8}$
\cite{oesbs}, and $\rm Sr_3Ru_2O_7$ \cite{rost:2011} is
displayed in Fig.~\ref{fig003}. This figure shows the main
features of the scaling behavior of the normalized effective
mass $M^*_N$ as given by Eq.~\eqref{UN2}. At low temperatures
$T_N<1$ the normalized effective mass is in the LFL domain; with
rising $T_N$ it crosses over through the transition, finally
entering the NFL regime.  The solid curve is the result of our
calculation of the scaling behavior. It is seen that $\rm
Sr_3Ru_2O_7$, located at the metamagnetic transition, and the
two HF metals exhibit the same scaling behavior, which is common
to other HF compounds and which can be understood within the
framework of fermion condensation or flat-band theory
\cite{ks91,pr,book,qp1,qp2}. As we will be emphasized below,
large temperature ranges are symptomatic of new quasiparticles,
and it is the scaling behavior of the normalized effective mass
that allows us to explain different properties of HF compounds
in their LFL, crossover, and NFL regimes and to demonstrate that
HF compounds, expressing universal scaling behavior, represent a
new state of matter.

\section{Strongly correlated quantum spin
liquids and quasicrystals} \label{QSL}

\subsection{Introduction}

In a frustrated magnet, spins are prevented from forming an
ordered alignment, so even at temperatures close to absolute
zero they collapse into a liquid-like state called a quantum
spin liquid (QSL). The herbertsmithite $\rm ZnCu_3(OH)_6Cl_2$
has been exposed as a $S=1/2$ kagome antiferromagnet
\cite{herb}, and recent experimental investigations have
revealed its unusual behavior \cite{helt,herb2,herb3}. Because
of its electrostatic environment, $\rm Cu^{2+}$ is expected to
occupy the distorted octahedral kagome sites. Magnetic kagome
planes $\rm Cu^{2+}$ $S=1/2$ are separated by nonmagnetic $\rm
Zn^{2+}$ layers. Observations have found no evidence of
long-range magnetic order or spin freezing down to temperature
of 50 mK, indicating that $\rm ZnCu_3(OH)_6Cl_2$ is the best
model found of the quantum kagome lattice
\cite{helt,herb2,herb3}. These results are confirmed by
theoretical considerations demonstrating that the ground state
of kagome antiferromagnet is a gapless spin liquid
\cite{shaginyan:2011,shaginyan:2012:A,shaginyan:2011:C,Normand}.
On the other hand, it has recently been suggested that there
exists a small spin-gap in the kagome layers
\cite{Han,Han11,sc_han}. The results reported are based on both
experimental facts and theoretical interpretation of these
results within the framework of impurity model. The experimental
facts are derived from high-resolution low-energy inelastic
neutron scattering on single-crystal $\rm ZnCu_{3}(OH)_6Cl_2$
Herbertsmithite, with the prospect of disentangling the effects
on the observed properties of this material due to $\rm Cu$
impurity spins from the effects of the kagome lattice itself
\cite{Han}. The impurity model assumes that the corresponding
impurity system may be represented as a simple cubic lattice in
the dilute limit below the percolation threshold.  The model
then suggests that the spin gap survives under the application
of magnetic fields up to 9 T \cite{sc_han}, while in the absence
of magnetic fields the bulk spin susceptibility $\chi$ exhibits
a divergent Curie-like tail, indicating that some of the $\rm
Cu$ spins act like weakly coupled impurities
\cite{Han,Han11,sc_han}. We will argue that the proposed
impurity model is artificial because it is impossible to isolate
the contributions coming from the impurities and the kagome
planes, in that the impurities and the kagome planes should be
considered as an integral system.  The model is therefore
inconsistent with the intrinsic properties of $\rm
ZnCu_{3}(OH)_6Cl_2$ as observed and described in recent
experimental and theoretical studies of the behavior of its
thermodynamic, dynamic, and relaxation properties. We
demonstrate that explanation of these properties lies in the
physics of the strongly correlated quantum spin liquid (SCQSL)
present in this system, for the behavior of $\rm
ZnCu_{3}(OH)_6Cl_2$ is in fact similar to that of heavy-fermion
metals and quasicrystals, with one main exception --- $\rm
ZnCu_{3}(OH)_6Cl_2$ does not support an electrical current
\cite{pr,shaginyan:2011,shaginyan:2012:A,shaginyan:2011:C,
shaginyan:2013:D,hfliq,comm,book,sc2016}. We conclude by
outlining a program to clarify the existence of SCQSL.  In
particular, we suggest that measurements of heat transport and
inelastic neutron scattering in magnetic fields $B$ be carried
out. Such measurements could be crucial in revealing both the
mechanisms involved and the real physics of QSL in $\rm
ZnCu_{3}(OH)_6Cl_2$.

\subsection{Thermodynamic properties}

\begin{figure} [! ht]
\begin{center}
\includegraphics [width=1.0\textwidth]{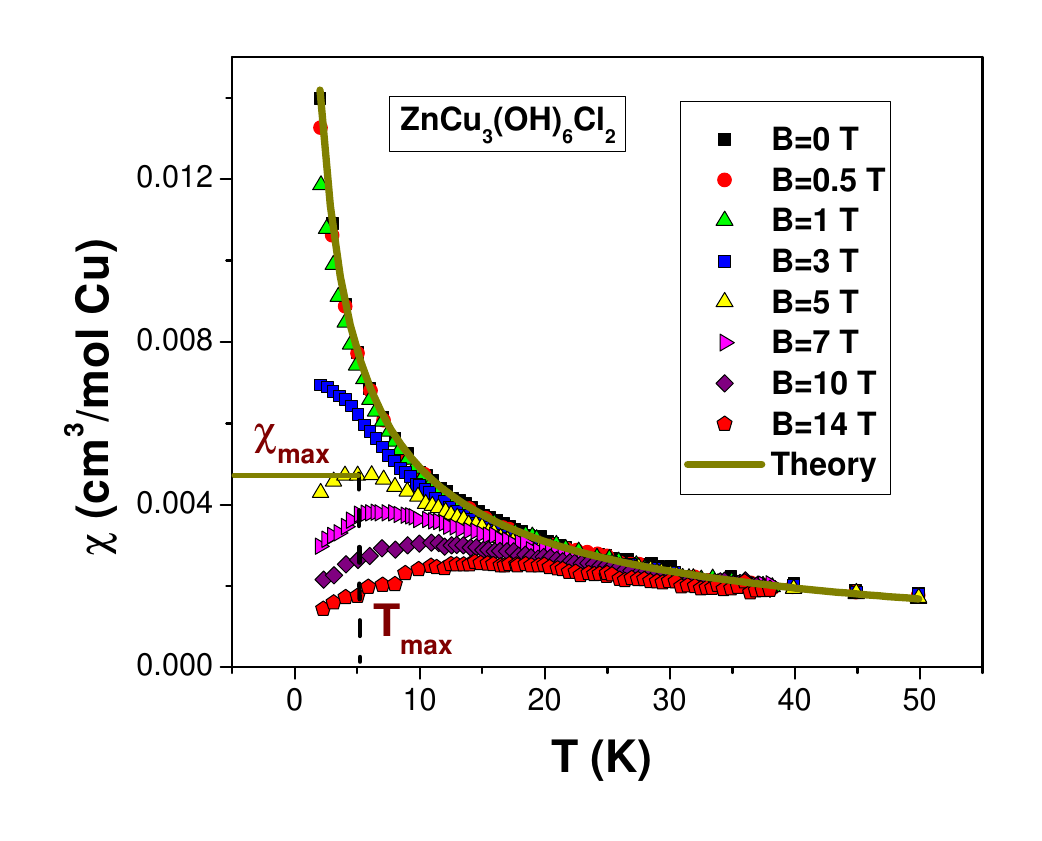}
\end{center}
\vspace*{-0.8cm} \caption{(Color online) Measured temperature
dependence of the magnetic susceptibility $\chi$ of $\rm
ZnCu_3(OH)_6Cl_2$ from Ref.~\cite{herb3} at magnetic fields
shown in the legend.  Illustrative values of $\chi_{\rm max}$
and $T_{\rm max}$ at $B=3$ T are also shown. A theoretical
prediction at $B=0$ is plotted as the solid curve, which
represents $\chi(T)\propto T^{-\alpha}$ with $\alpha=2/3$
\cite{shaginyan:2011,book}.} \label{fig01}
\end{figure}

\begin{figure} [! ht]
\begin{center}
\includegraphics [width=1.0\textwidth]{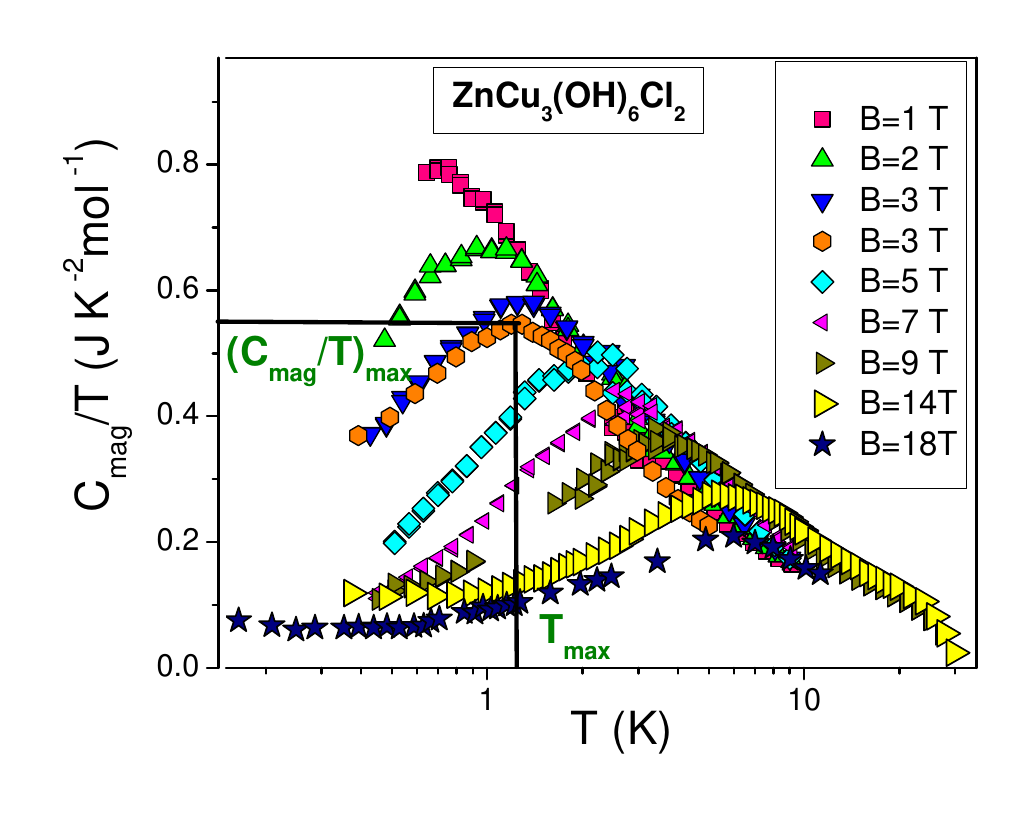}
\end{center}
\vspace*{-0.8cm} \caption{(Color online) Specific heat $C_{\rm
mag}/T$ measured on powder \cite{helt,herb2} and single-crystal
\cite{herb,t_han:2012,t_han:2014} samples of Herbertsmithite is
displayed as a function of temperature $T$ for fields $B$ shown
in the legend.} \label{fig02}
\end{figure}

\begin{figure} [! ht]
\begin{center}
\includegraphics [width=1.0\textwidth]{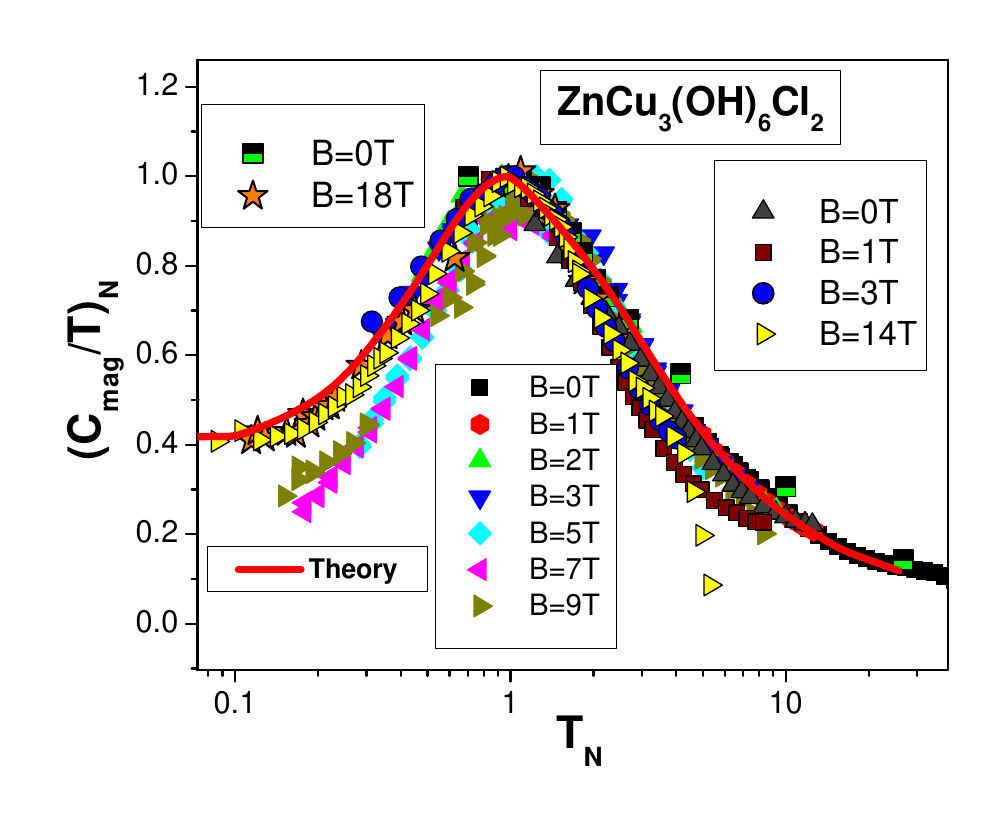}
\end{center}
\vspace*{-0.8cm} \caption{(Color online) Normalized specific
heat $(C_{\rm mag}/T)_N$ versus normalized temperature $T_N$ at
$B$-field values shown in the legend \cite{hfliq,book}. The
theoretical result from Refs.~\cite{shaginyan:2011,book},
represented by the solid curve, traces the scaling behavior of
the effective mass.} \label{fig03}
\end{figure}

To examine the impurity model in a broader context, we first
refer to the experimental behavior of the magnetic
susceptibility $\chi$ of Herbertsmithite.  It is seen from
Fig.~\ref{fig01} that the magnetic susceptibility has the
divergent behavior $\chi(T)\propto T^{-2/3}$, in magnetic fields
$B\leq 1$ T, as shown by the solid line.  In the case of weakly
interacting impurities it is suggested that the low-temperature
behavior of $\chi_{\rm CW}(T)\propto 1/(T+\theta)$ can be
approximated by a Curie-Weiss law \cite{Han,Han11,sc_han}, with
$\theta$ a vanishingly small Weiss temperature. However, given
that $\chi(T)\propto T^{-2/3}$, the Curie-Weiss approximation is
in conflict with both experiment \cite{herb3} and theory
\cite{shaginyan:2011,book}. Within the framework of the impurity
model, the calculated intrinsic spin susceptibility of the
kagome plane obeys $\chi_{\rm kag}(T)=\chi(T) -\chi_{\rm
CW}(T)$, leading to $\chi_{\rm kag}(T\to0)\to 0$ and the
erroneous claim that a putative gap has been observed
\cite{Han11}. Thus, we must conclude that the impurity model is
not valid, since it cannot explain the empirically validated
behavior $\chi(T)\propto T^{-2/3}$. To explain this behavior of
$\chi$, it is necessary to consider the impurities and the
kagome planes as an integral system
\cite{pr,shaginyan:2011,shaginyan:2012:A,shaginyan:2011:C,shaginyan:2013:D,comm,hfliq,book,sc2016}.

In similar vein, working within the impurity model the authors
of Ref.~\cite{Han} obtain a measure
$S_{kag}(\omega)=S_{tot}(\omega) -aS_{\rm imp}(\omega)$ of the
intrinsic scattering by subtracting the impurity scattering
$S_{\rm imp}(\omega)$ from the total scattering
$S_{tot}(\omega)$, taking $a$ as a fitting parameter. As a
result, they find that $S_{\rm kag}(\omega)\to 0$ as $\omega$
decreases below an energy of 0.7 meV (see Fig.~4(b) of
Ref.~\cite{Han}) and claim observation of a gap, whereas we have
shown above that such a subtraction leads to the erroneous
conclusion that a gap has been found.  Indeed, this conclusion
relies completely on the theoretical assumption that the
impurities are weakly interacting; accordingly, it cannot be
considered as experimental fact.

Let us consider somewhat further the inadequacy of the impurity
model and its corresponding gap, when confronted with
experimental findings.  We see from Fig.~\ref{fig01} that Landau
Fermi liquid (LFL) behavior is demonstrated at least for $B\geq
3$ T and low temperatures $T$. At such temperatures and magnetic
fields the impurities should become fully polarized.  They do
not exhibit typical Curie-Weiss behavior; otherwise the impurity
mode would fail. Thus, assuming the impurities are fully
polarized and hence do do not contribute to $\chi$, one has
simply $\chi_{\rm kag}(T) =\chi(T)$. Analogous behavior for the
heat capacity follows from Fig.~\ref{fig02}. LFL behavior of
$C_{\rm mag}/T$ emerges under application of the same fields.
Consequently, we may conclude that at least at $B\geq 3$ T and
low $T$, the contributions to both $\chi$ and $C_{\rm mag}/T$
from the impurities are negligible; rather, one expects them to
be dominated by the kagome lattice, exhibiting a spin gap in the
kagome layers \cite{Han,Han11,sc_han}. Thus, according to the
impurity model one would expect both $\chi(T)$ and $C_{\rm
mag}(T)/T$ to approach zero for $T \to 0$ at $B\geq 3$ T. From
Figs.~\ref{fig01}, \ref{fig02}, and \ref{fig03}, it is clear
that this is not the case.  Up to $B\sim 14$ T neither $\chi$
nor $C_{\rm mag}/T$ approaches zero as $T\to0$. Moreover, the
normalized $C_{\rm mag}/T$ follows the uniform scaling behavior
displayed in Fig.~\ref{fig03}, confirming the absence of a gap.
Also, as indicated in Fig.~\ref{fig02}, it is found that the
recent measurements of $C_{\rm mag}$
\cite{helt,herb2,herb,t_han:2012,t_han:2014} are compatible with
those obtained on powder samples. All relevant experimental
observations support the conclusions that (i) the properties of
$\rm ZnCu_3(OH)_6Cl_2$ under study are determined by a stable
SCQSL, (ii) there is no appreciable gap in the spectra of spinon
excitations, such a gap being absent even under the application
of very high magnetic fields of 18 T, and (iii) the impurity
model is untenable from the experimental standpoint.  These
conclusions agree with recent experimental findings that the
low-temperature plateau in local susceptibility identifies the
spin-liquid ground state as being gapless \cite{zorko}, while
recent theoretical analysis confirms the absence of a gap
\cite{Normand}.

\subsection{Relaxation and transport properties}

\begin{figure} [! ht]
\begin{center}
\includegraphics [width=1.0\textwidth]{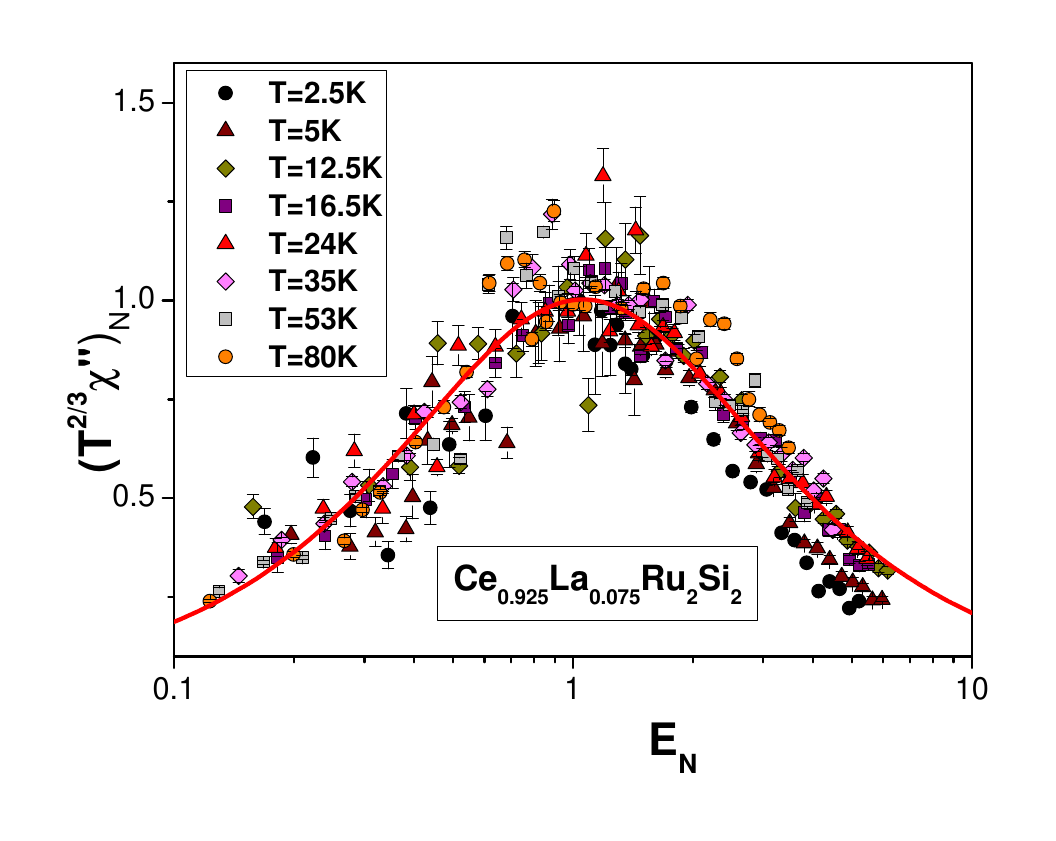}
\end{center}
\caption{(Color online) Scaling behavior of the normalized
dynamic spin susceptibility $(T^{2/3}\chi'')_N$. Data are
extracted from measurements on the heavy-fermion metal $\rm
Ce_{0.925}La_{0.075}Ru_2Si_2$ \cite{knafo:2004} and plotted
against the dimensionless variable $E_N$. Solid curve:
Theoretical calculations based on Eq.~\eqref{SCHIN}
\cite{shaginyan:2012:A}.}\label{fig04}
\end{figure}

\begin{figure} [! ht]
\begin{center}
\includegraphics [width=1.0\textwidth]{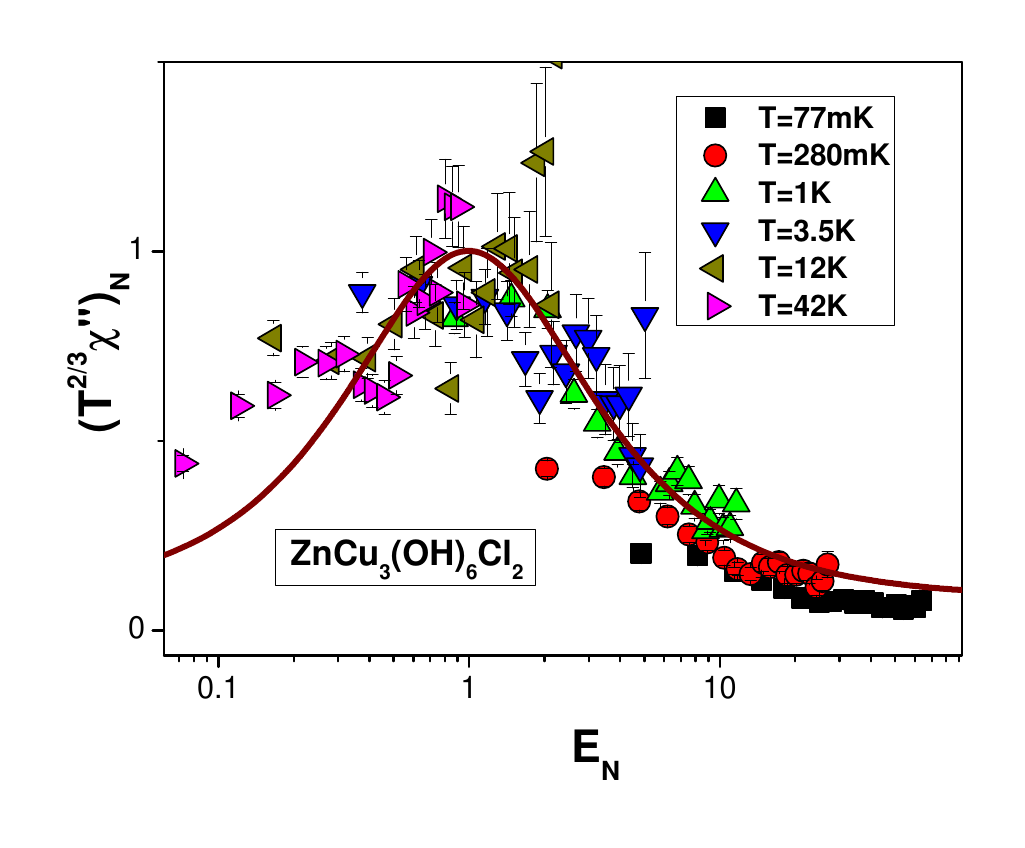}
\end{center}
\caption{(Color online) Scaling behavior of the normalized
dynamic spin susceptibility $(T^{2/3}\chi'')_N$. Data are
extracted from measurements on Herbertsmithite $\rm
ZnCu_3(OH)_6Cl_2$ \cite{herb3}. Solid curve: Theoretical
calculations based on Eq.~\eqref{SCHIN}
\cite{shaginyan:2012:A}.}\label{fig05}
\end{figure}

\begin{figure} [! ht]
\begin{center}
\includegraphics [width=1.0\textwidth]{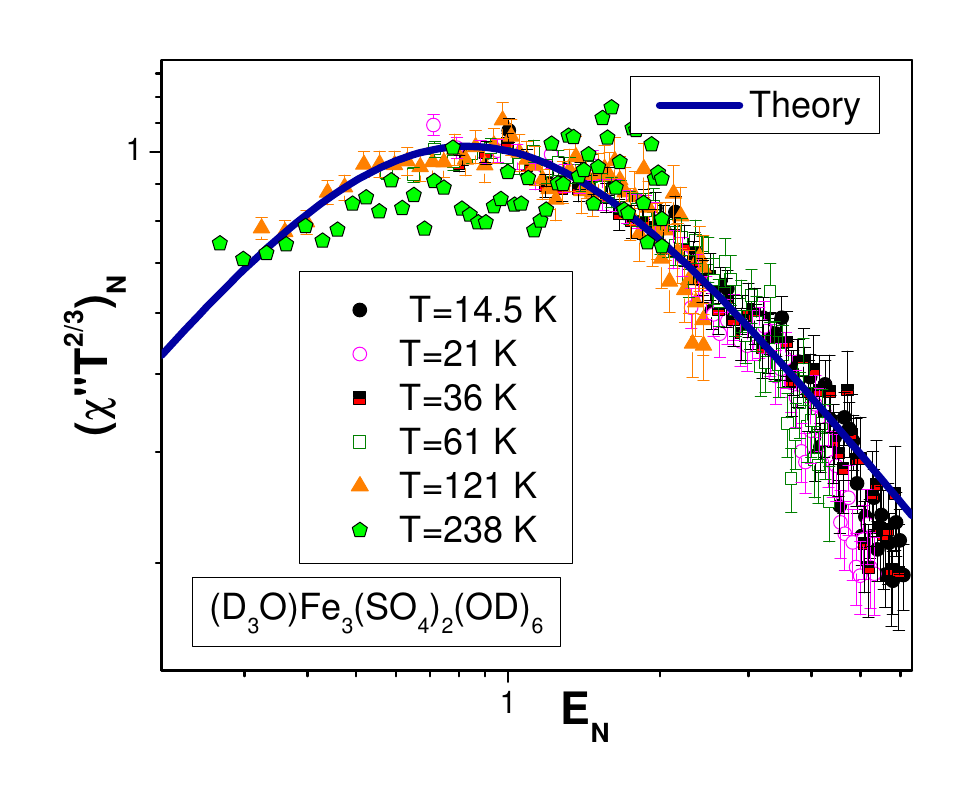}
\end{center}
\caption{(Color online) Scaling behavior of the normalized
dynamic spin susceptibility $(T^{2/3}\chi'')_N$. Data are
extracted from measurements on the deuteronium jarosite $\rm
(D_3O)Fe_3(SO_4)_2(OD)_6$ \cite{faak:2008}. Solid curve reports
the theoretical calculations based on Eq.~\eqref{SCHIN}
\cite{shaginyan:2012:A}.}\label{fig06}
\end{figure}

\begin{figure} [! ht]
\begin{center}
\includegraphics [width=1.0\textwidth]{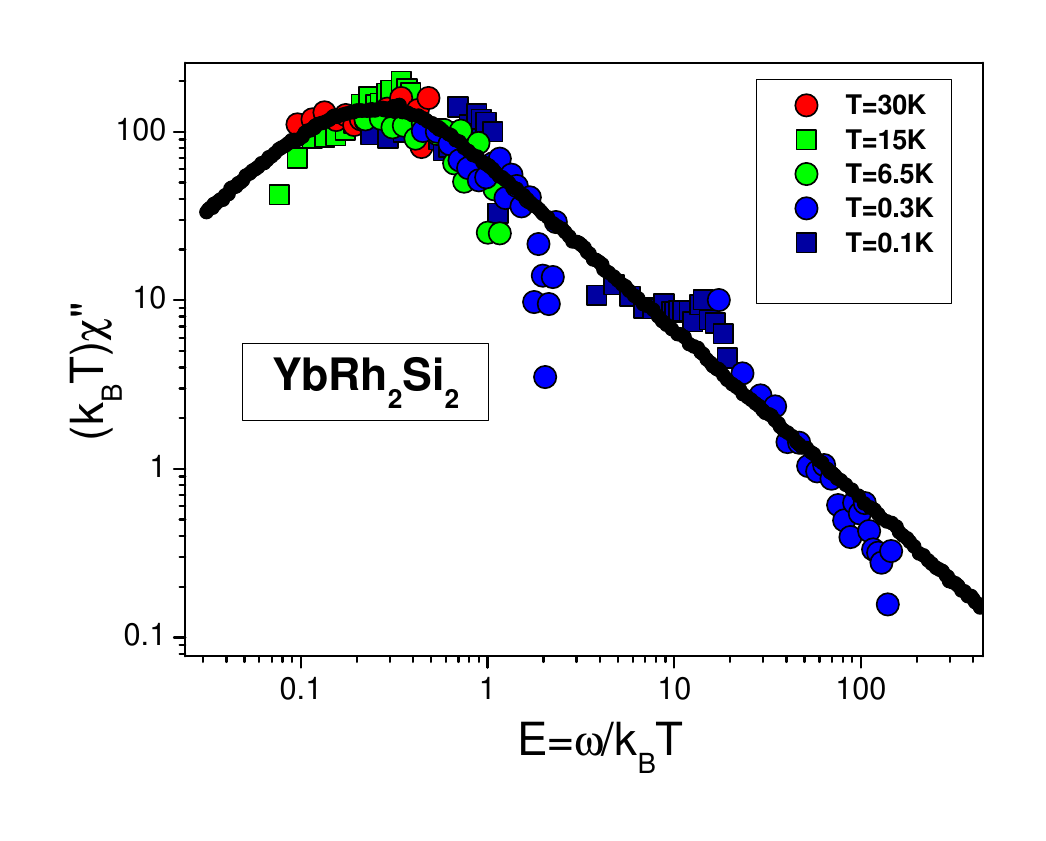}
\end{center}
\caption{(Color online) Scaling behavior of the normalized
dynamic spin susceptibility $T\chi''$ plotted against
$E=\omega/k_BT$. The data are extracted from measurements on
$\rm YbRh_2Si_2$ \cite{stock}. The solid curve is fitted with
the function given by Eq.~\eqref{SCHIT}.}\label{fig09}
\end{figure}

The same conclusions can be drawn from the results of
neutron-scattering measurements of the dynamic spin
susceptibility $\chi({\bf q},\omega,T) =\chi{'}({\bf
q},\omega,T)+i\chi{''}({\bf q},\omega,T)$ as a function of
momentum $q$, frequency $\omega$, and temperature $T$. Indeed,
these results play a crucial role in identifying the properties
of the quasiparticle excitations involved. At low temperatures,
such measurements reveal that the corresponding quasiparticles
-- of a new type insulator -- are represented by spinons, form a
continuum, and populate an approximately flat band crossing the
Fermi level \cite{Han:2012}.  In such a situation it is expected
that the dimensionless normalized susceptibility
$(T^{2/3}\chi'')_{N}=T^{2/3}\chi'' /(T^{2/3}\chi'')_{\rm max}$
exhibits scaling as a function of the dimensionless energy
variable $E_N = E/E_{\rm max}$ \cite{shaginyan:2012:A,book}.
Specifically, the equation describing the normalized
susceptibility $(T^{2/3}\chi'')_{N}$ reads
\cite{shaginyan:2012:A,book}
\begin{equation}\label{SCHIN}
(T^{2/3}\chi'')_N\simeq\frac{b_1E_N}{1+b_2E_N^2},
\end{equation}
where $b_1$ and $b_2$ are fitting parameters adjusted such that
the function $(T^{2/3}\chi'')_{N}$ reaches its maximum value
unity at $E_N=1$ \cite{book,shaginyan:2012:A}.
Figure~\ref{fig04} displays  $(T^{2/3}\chi'')_{N}$ values
extracted from measurements of the inelastic neutron-scattering
spectrum on the heavy-fermion (HF) metal $\rm
Ce_{0.925}La_{0.075}Ru_2Si_2$ \cite{knafo:2004}. The scaled data
for this quantity, obtained from measurements on two quite
different strongly correlated systems, $\rm ZnCu_3(OH)_6Cl_2$
\cite{herb3} and $\rm (D_3O)Fe_3(SO_4)_2(OD)_6$
\cite{faak:2008}, are displayed in Figs.~\ref{fig05} and
\ref{fig06}, respectively. It is seen that the theoretical
results from Ref.~\cite{shaginyan:2012:A} (solid curves) are in
good agreement with the experimental data collected on all three
compounds over almost three orders of magnitude of the scaled
variable $E_N$.  Hence $(T^{2/3}\chi'')_{N}$ does exhibit the
anticipated scaling behavior for these systems. From this
observation we infer that the spin excitations in both $\rm
ZnCu_3(OH)_6Cl_2$ and $\rm (D_3O)Fe_3(SO_4)_2(OD)_6$ demonstrate
the same itinerate behavior as the electronic excitations of the
HF metal $\rm Ce_{0.925}La_{0.075}Ru_2Si_2$ and therefore form a
continuum. This detection of a continuum is of great importance
since it clearly signals the presence of a SCQSL in
Herbertsmithite \cite{shaginyan:2012:A,shaginyan:2011:C,book}.

Provided that a fermion condensate (FC) is indeed present in the
electronic system of a HF metal, we know that the imaginary part
$\chi''(T,\omega)$ of the susceptibility is given by
\cite{book,JETP}
\begin{equation}\label{SCHIT}
T\chi''(T,\omega)\simeq\frac{a_5E}{1+a_6E^2},
\end{equation}
where $E=\omega/k_BT$ while $a_5$ and $a_6$ are constants. It is
seen from Eq.~\eqref{SCHIT} that $T\chi''(T,\omega)$ depends on
the only the variable $E=\omega/k_BT$. Thus, Eqs.~\eqref{SCHIN}
and \eqref{SCHIT} establish two types of scaling behavior of
$\chi''(\omega,T)$. In Fig.~\ref{fig09}, the dynamic
susceptibility $(T\chi'')$ extracted from measurements of the
inelastic neutron scattering spectrum on the HF metal $\rm
YbRh_2Si_2$ \cite{stock} is shown. The data for $(T\chi'')$
exhibit scaling behavior over three decades in the variation of
both this function and the variable $E$, thus confirming the
validity of Eq.~\eqref{SCHIT}. The scaled data obtained in
measurements on such quite different strongly correlated systems
as $\rm ZnCu_3(OH)_6Cl_2$, $\rm Ce_{0.925}La_{0.075}Ru_2Si_2$,
$\rm (D_3O)Fe_3(SO_4)_2(OD)_6$, and $\rm YbRh_2Si_2$ collapse
fairly well onto a single curve over almost three decades of the
scaled variables.

It is apparent from Figs.~\ref{fig04}, \ref{fig05}, \ref{fig06},
and \ref{fig09} that the calculations based on this premise are
in good agreement with the experimental data, affirming the
identification of SCQSL as the agent responsible for the
low-temperature behavior of $\rm ZnCu_3(OH)_6Cl_2$ and $\rm
(D_3O)Fe_3(SO_4)_2(OD)_6$. We conclude that the concept of the
spin gap in the kagome layers is an artificial construct at odds
with known properties of $\rm ZnCu_{3}(OH)_6Cl_2$, thereby
negating the existence of a spin gap in the SCQSL of
Herbertsmithite. It is noted that the presence of a gap in the
kagome layers is not in itself of vital importance, for it does
not govern the thermodynamic and transport properties of $\rm
ZnCu_3(OH)_6Cl_2$. Rather, these properties are determined by
the underlying SCQSL. This assertion can be tested by
measurements of the heat transport in magnetic fields, as has
been done successfully in the case of the organic insulators
$\rm EtMe_3Sb[Pd(dmit)_2]_2$ and $\rm
\kappa-(BEDT-TTF)_2Cu_2(CN)_3$
\cite{yamashita:2010,yamashita:2012,shaginyan:2013:D}.
Measurements of thermal transport are particularly salient in
that they probe the low-lying elementary excitations of SCQSL in
$\rm ZnCu_3(OH)_6Cl_2$ and potentially reveal itinerant spin
excitations that are mainly responsible for the heat transport.
Surely, the overall heat transport is contaminated by the phonon
contribution; however, this contribution is hardly affected by
the magnetic field $B$. Essentially, we expect that measurement
of the $B$-dependence of thermal transport will be an important
step toward resolving the nature of the SCQSL in $\rm
ZnCu_3(OH)_6Cl_2$ \cite{shaginyan:2011:C,shaginyan:2013:D,book}.

The SCQSL in Herbertsmithite behaves like the electron liquid in
HF metals -- provided the charge of an electron is set to zero.
As a result, the thermal resistivity $w$ of the SCQSL is given
by \cite{shaginyan:2011:C,shaginyan:2013:D,book}
\begin{equation}\label{wr}
w-w_0=W_rT^2\propto\rho-\rho_0\propto(M^*)^2T^2,
\end{equation}
where $W_r{T^{2}}$ represents the contribution of spinon-spinon
scattering to thermal transport, being analogous to the
contribution $AT^2$ to charge transport from electron-electron
scattering. Here $\rho$ is the longitudinal magnetoresistivity
(LMR), and $w_0$ and $\rho_0$ are the residual thermal
resistivity and residual resistivity, respectively.

\begin{figure} [! ht]
\begin{center}
\includegraphics [width=1.0\textwidth]{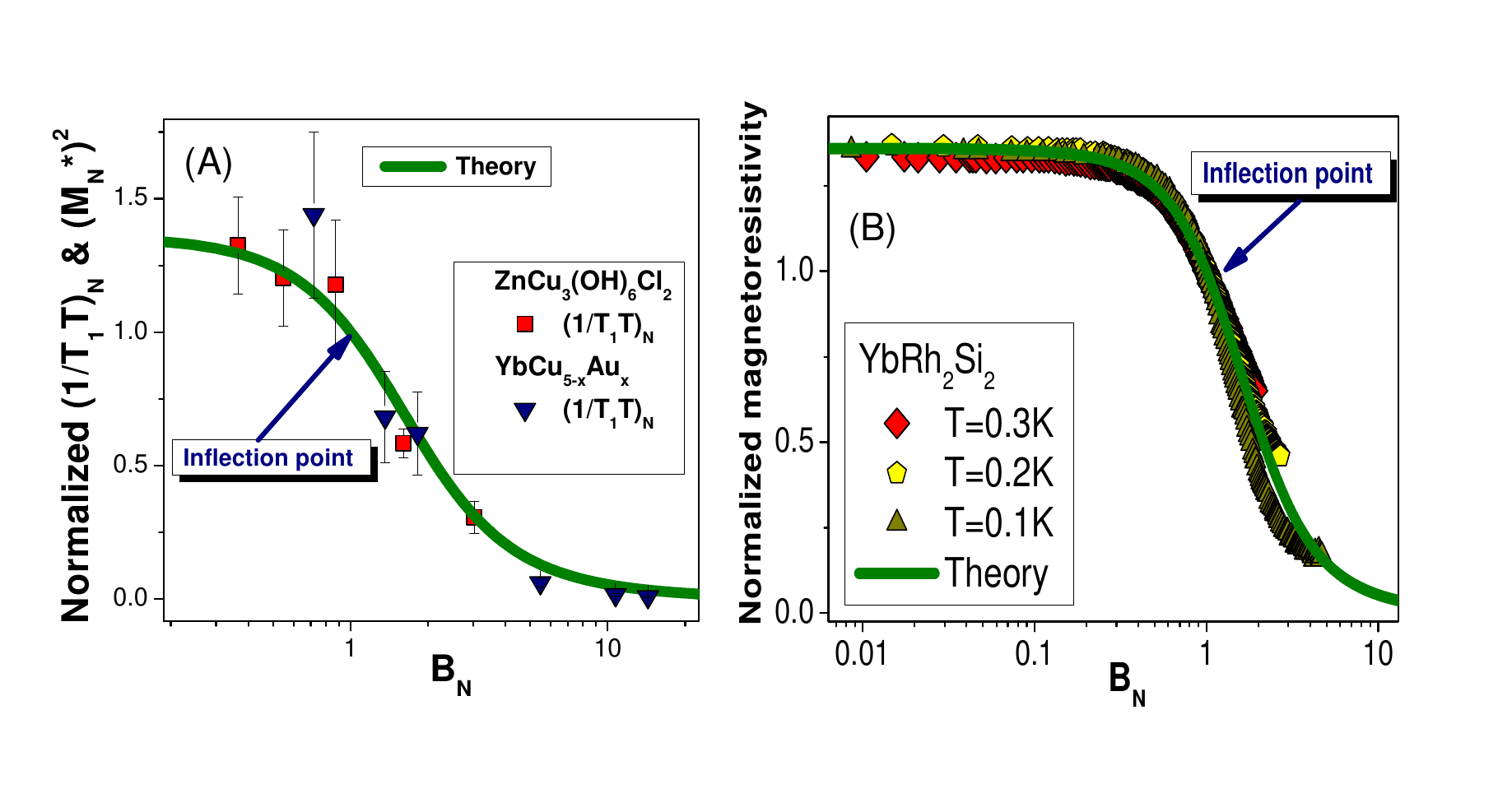}
\end{center}
\caption{(Color online) Panel (A). Normalized spin-lattice
relaxation rate $(1/T_1T)_N$ at fixed temperature as a function
of magnetic field. Data for $(1/T_1T)_N$ extracted from
measurements on $\rm ZnCu_3(OH)_6Cl_2$ are shown by solid
squares \cite{imai} and those extracted from measurements on
$\rm YbCu_{5-x}Au_{x}$ at $x=0.4$, by the solid triangles
\cite{carr}. The inflection point at which the normalization is
taken is indicated by the arrow. Panel (B). Magnetic field
dependence of the normalized magnetoresistance $\rho_N$,
extracted from LMR of $\rm YbRh_2Si_2$ at different temperatures
\cite{gegmr} listed in the legend. The inflection point is shown
by the arrow. In both panels (A) and (B), the calculated result
is depicted by the same solid curve, tracing the scaling
behavior of $W_r\propto(M^*)^2$ (see
Eq.~\eqref{WT}).}\label{T12}
\end{figure}

\begin{figure} [! ht]
\begin{center}
\includegraphics [width=1.0\textwidth]{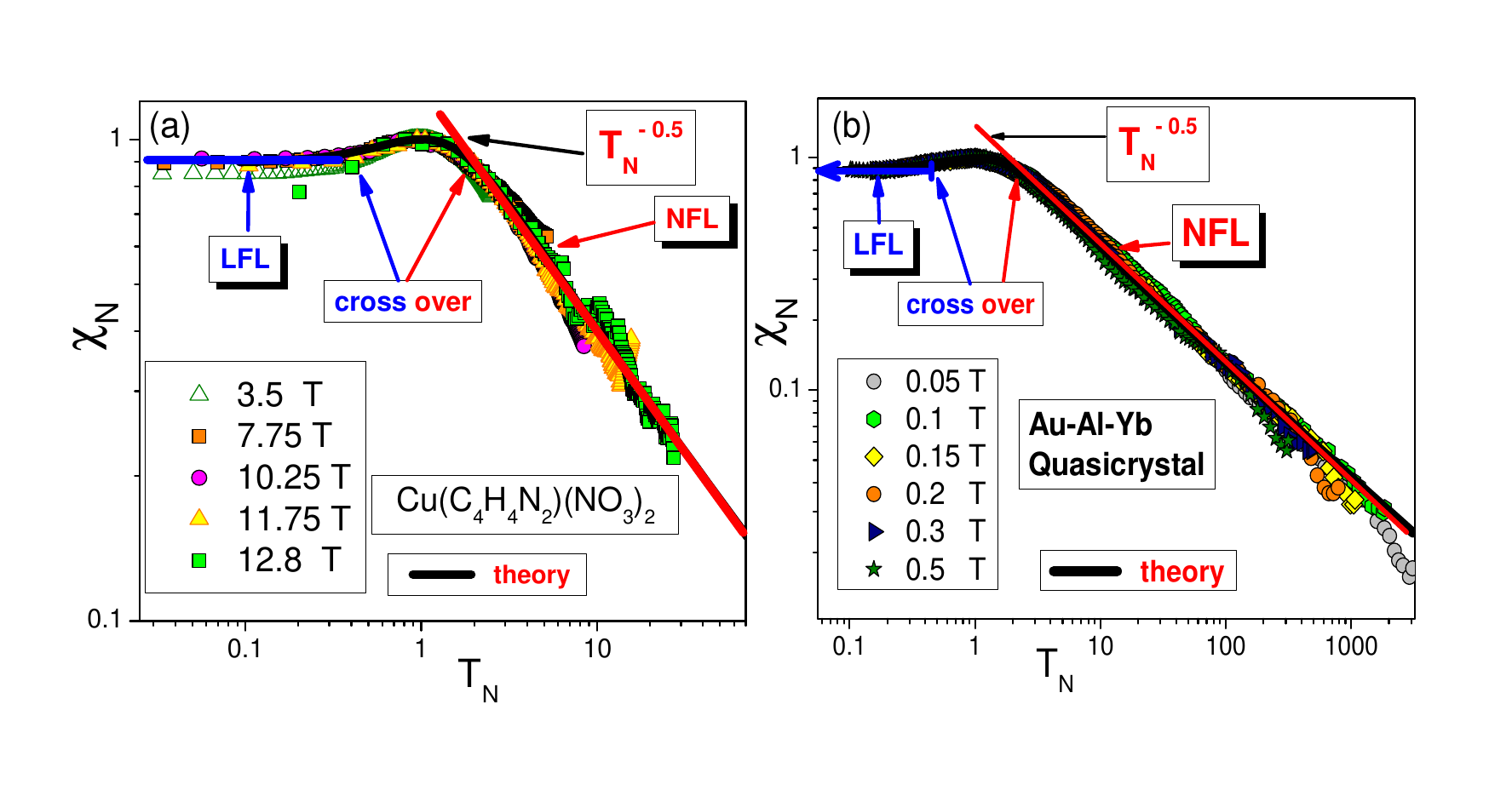}
\end{center}
\caption{(Color online) Normalized magnetic susceptibility
$\chi_N$ extracted from measurements in magnetic fields $H$
(shown in the legend) on CuPzN \cite{prl15} (panel (a)) and on
the Au$_{51}$ Al$_{34}$Yb$_{15}$ quasicrystal \cite{QCM} (panel
(b)). Our corresponding theoretical curves, merged in the scale
of the figure, are drawn as the solid lines tracing the scaling
behavior.  Panels (a) and (b) demonstrate that the dependence
$\chi_N$ versus $T_N$ for CuPzN and for the quasicrystal has
three distinctive regions: LFL, crossover, and NFL, showing
behavior $\chi_N\sim T_N^{-0.5}$ in the latter regime (orange
curve).}\label{fig08}
\end{figure}

\begin{figure} [! ht]
\begin{center}
\includegraphics [width=1.0\textwidth]{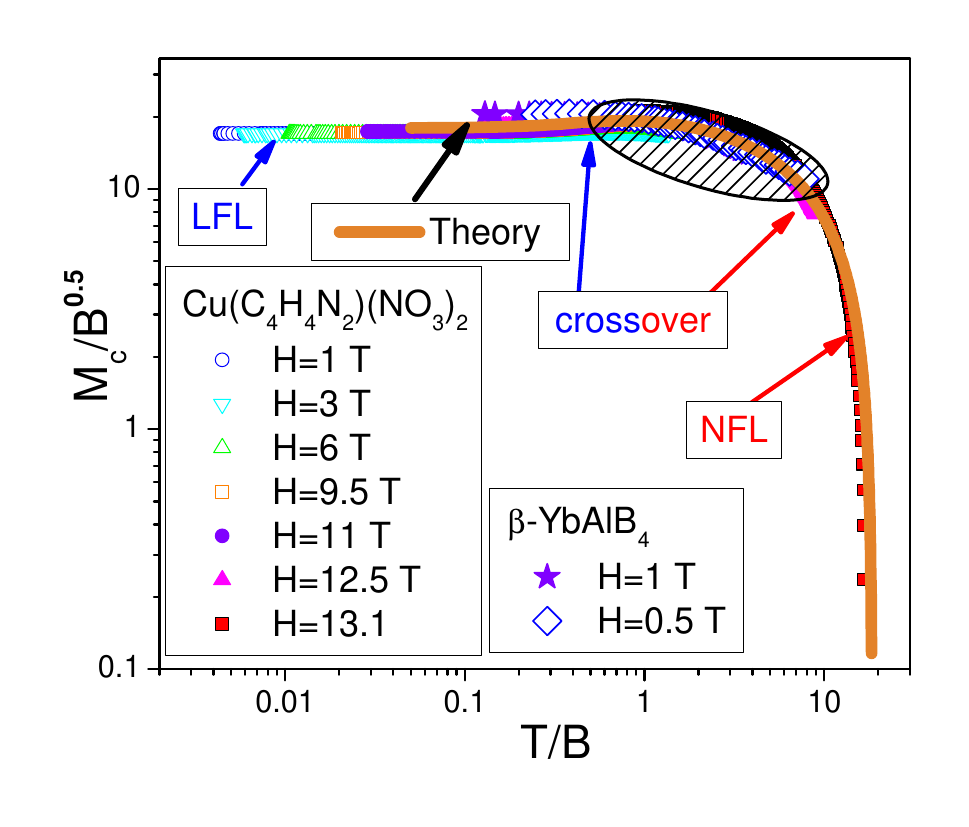}
\end{center}
\caption{(Color online) Scaling dependence of the function
$M_c/\sqrt{B}$ ($M_c=M_s-M$, $B=H_s-H$) on $(T/B)_N$ for CuPzN.
Experimental data are taken from Refs.~
\cite{prl15,s2011,s2015}. The curves correspond to different
magnetic fields $H$ listed in the legends. LFL, crossover, and
NFL regions are shown. The theoretical dependence is represented
by the solid curve \cite{annphys}.}\label{fig10}
\end{figure}

We next consider the effect of a magnetic field $B$ on the
spin-lattice relaxation rate {$1/(T_1T)$}. Referring to panel
(A) of Fig.~\ref{T12}, which shows the normalized spin-lattice
relaxation rate $1/(T_1T)_N$ at fixed temperature versus
magnetic field $B$, it is seen that increasing $B$ progressively
reduces { $1/(T_1T)$}, and that as a function of $B$, there is
an inflection point at some $B=B_{\rm inf}$, marked by the
arrow. To clarify the scaling behavior in this case, we
normalize $1/(T_1T)$ by its value at the inflection point, while
the magnetic field is normalized by $B_{\rm inf}$.  Taking into
account the relation $1/(T_1T)_N\propto(M^*)^2$, we expect that
a strongly correlated Fermi system located near its quantum
critical point will exhibit behavior similar to { $1/(T_1T)_N$}
\cite{pr,shaginyan:2011:C,shaginyan:2013:D,book}. Based on this
reasoning, it follows that with application of magnetic fields
at fixed temperature, the coefficient $W_r$ behaves like the
spin-lattice relaxation rate shown in Fig.~\ref{T12}, i.e.,
$W_r\propto 1/(T_1T)$
\cite{shaginyan:2011:C,shaginyan:2013:D,book}. Significantly,
panel A of Fig.~\ref{T12} shows that the Herbertsmithite $\rm
ZnCu_3(OH)_6Cl_2$ \cite{imai} and the HF metal $\rm
YbCu_{5-x}Au_{x}$ \cite{carr} do in fact show the same behavior
for the normalized spin-lattice relaxation rate. As indicated in
Fig.~\ref{T12}, for $B\leq B_{\rm inf}$ (or $B_N\leq1$) the
normalized relaxation rate $1/(T_1T)_N$ depends weakly on the
magnetic field, while it diminishes at the higher fields
\cite{pr,shaginyan:2011:C,shaginyan:2013:D,book}. Panel B of
Fig. \ref{T12} reports results for the normalized
magnetoresistance
\begin{equation}\label{rn06}
\rho_N(y)\equiv\frac{\rho(y)-\rho_0}{\rho_{inf}}=(M_N^*(y))^2
\end{equation}
versus normalized magnetic field $y=B/B_{inf}$ at the three
temperatures shown in the legend. Here $\rho_{inf}$ and
$B_{inf}$ are the longitudinal magnetoresistance (LMR) and
magnetic field, respectively, taken at the inflection point
indicated by the arrow in panel B of Fig.~\ref{T12}. Both
theoretical (solid line) and experimental (symbols) curves have
been normalized by their inflection points, which also reveal
the scaling behavior: The scaled curves at different
temperatures merge into a single curve as a function of the
variable $B/B_{inf}$ to establish scaling behavior over three
orders in this normalized magnetic field. Taking into account
Eq. \eqref{wr}, we obtain
\begin{equation}\label{WT}
W_r\propto{ 1/(T_1T)_N}\simeq \rho_N\simeq(M^*)_N^2\propto
B^{-4/3}.
\end{equation}
We thus predict that the thermal resistivity of $\rm
ZnCu_3(OH)_6Cl_2$ behaves like the magnetoresistance of the
archetypical HF metal $\rm YbRh_2Si_2$, and we conclude that
application of a magnetic field $B$ leads to a crossover from
NFL to LFL behavior and to a significant reduction in both the
relaxation rate and the thermal resistivity.  We note that in
order to directly observe a possible gap, it will also be
crucial to carry out measurements of low-energy inelastic
neutron scattering on single crystals of $\rm ZnCu_3(OH)_6Cl_2$
subject to a magnetic field that drives the system toward LFL
behavior, since in that case the contribution coming from
supposed impurities is negligible, as addressed above for the
spin susceptibility $\chi$.

\subsection{One-dimensional spin liquids and quasicrystals}

To compare the scaling behavior of one-dimensional (1D) SCQSL
with that found in other HF compounds such as the quasicrystal
Au$_{51}$Al$_{34}$Yb$_{15}$, we turn to Fig.~\ref{fig08}, which
portrays the comparison between the normalized susceptibility
$\chi_N$ extracted from experiments on the insulator $\rm
Cu(C_4H_4N_2)(NO_3)_2$ (CuPzN), holding 1D spin liquid,
\cite{prl15} (panel (a)), Au$_{51}$Al$_{34}$Yb$_{15}$ \cite{QCM}
(panel (b)), and the theory developed for quasicrystals and 1D
SCQSL \cite{quasicryst,annphys}.  For more then three decades in
normalized temperature there is very good agreement between the
theory and the experimental data. The double-log scale used for
the plots reveals the universal dependence $\chi_N \sim
T_N^{-0.5}$. Comparison of panels (a) and (b) indicates that in
both CuPzN and Au$_{51}$Al$_{34}$Yb$_{15}$, the normalized
susceptibility $\chi_N$ of the 1D SCQSL has three regions of
characteristic behavior: the low-temperature LFL part, the
medium-temperature crossover region where the maximum occurs,
and the high-temperature NFL part with the distinctive
temperature dependence $T_N^{-0.5}$.  The absolute values of the
thermodynamic functions obviously depend on the properties of
the individual system in question.  The universal features and
behavior shared by these systems have only been revealed by
means of the normalization procedure based on internal scales
\cite{quasicryst,pr,book}.

As a final exercise, we trace the scaling behavior of the
function $(M_s-M)/\sqrt{B}$ with respect to $T/B$, applied $B=
H_s-H$, where $H$ is the applied magnetic field while $M_s$ and
$H_s$ denote respectively the saturation magnetization and
saturation magnetic field \cite{prl15,s2011,s2015,annphys}. The
corresponding theoretical dependence can be inferred from
Eqs.~\eqref{UN2} and \eqref{NORM} exploiting the fact that the
magnetization is related to the susceptibility $\chi$ via
$M=\int \chi dH$, where $\chi \sim M^*(T,H)$ is the magnetic
susceptibility. Then, having the effective mass $M^*(T,H)$ from
the interpolative solution \eqref{UN2} of the Landau equation,
and setting $M_s-M = M_c$, we may obtain the corresponding
theoretical dependence \cite{annphys}
\begin{equation}\label{scmc}
\frac{M_c}{\sqrt{B}}=a+\frac{M_s-M}{\sqrt{H_s-H}},
\end{equation}
as a function of $T/B=T/(H_s-H)$.  That this result is in good
agreement with the experimental facts may be seen from
Fig.~\ref{fig10} which illustrates the scaling behavior of the
magnetization $M_c$. This behavior, expressed by
Eq.~\eqref{scmc} and tracked by the solid curve, is
indistinguishable from the dependence $M_c/B^{0.5}$ versus $T/B$
extracted from the experimental data \cite{prl15,s2011,s2015}, a
remarkable coincidence. According to Fig.~\ref{fig10}, LFL
regime occurs at $T<<B$, the crossover around $T \sim B$, and
the NFL regime $T>>B$, this being the case for HF compounds that
behave like the HF superconductor $\beta$ - YbAlB$_4$, the
quasicrystal $\rm Au_{51}Al_{34}Yb_{15}$, the SCQSL of $\rm
ZnCu_3(OH)_6Cl_2$, and the 1D SCQSL of CuPzN
\cite{quasicryst,pr,book,annphys}.

\subsection{Summary}

To summarize the essence of this section, we have demonstrated
that both the impurity model of Herbertsmithite $\rm
ZnCu_{3}(OH)_6Cl_2$ and the existence of a spin gap in this
compound are problematic, as they contradict established
properties of this system and are not supported by detailed
considerations of its thermodynamics and relaxation properties
in magnetic fields.  In conclusion, we recommend that
measurements of heat transport in magnetic fields be conducted
to clarify the nature of the quantum spin-liquid this system. We
also have suggested that measurements of low-energy inelastic
neutron scattering on $\rm ZnCu_3(OH)_6Cl_2$ single crystals be
performed in magnetic fields.  In such a study the contribution
coming from supposed impurities would be negligible.  We have
also shown that the strongly correlated quantum spin liquids
existing in $\rm ZnCu_{3}(OH)_6Cl_2$ and CuPzN exhibit the same
universal behavior as that of other HF compounds, thus providing
empirical evidence for a new state of matter related to the
presence of flat bands.

\section{Precursors of fermion condensation in a gas of 2D
ultra cold fermionic atoms}\label{gas}

Here we propose a simple model that describes the appearance of
precursors (at low but finite temperatures) of fermion
condensation (FC) \cite{book}  in two-dimensional ensembles of
ultracold fermionic atoms, interacting with coherent resonant
light. Latter interaction permits to introduce spin-orbit-like
and  Zeeman - like atom couplings, which at low temperatures
drive the system to FCQPT. We note that FC can take place in
finite systems as well \cite{pr,book,qp1}. We obtain the system
phase diagram in the $H-\gamma$ ($H$ is above Zeeman - like
field and $\gamma$ is a strength of interatomic interaction)
variables. We show that thermodynamic (magnetic moment and spin
susceptibility) and spectroscopic (photon absorption spectra)
characteristics of the system exhibit peculiar features due to
FC precursor realization. These features can be regarded as FC
fingerprints in the system under consideration. Recent progress
in achieving  highly coherent light-atom interactions in cold
atomic matter allows researchers to realize new quantum degrees
of freedom such as atomic pseudospin. This pseudospin is related
to the coupling of light to the hyperfine structure of the
atomic spectrum and can be realized even for bosonic atoms
having zero total physical spin like $^{87}$Rb isotope. Since
the atoms coupling to coherent highly resonant light strongly
depends on the atom velocity, the atomic motion generates a
linear (in the atomic momentum) pseudospin - momentum
interaction. This effect produces an artificial analog of the
spin-orbit coupling, linear both in the atomic momentum and
pseudospin. This interaction formally resembles the famous
spin-orbit coupling of electric charge carriers in solids. One
can also produce an artificial magnetic field acting on the
atomic pseudospin and corresponding to a strong Zeeman - like
interaction. These studies have been comprehensively reviewed in
Refs. \cite{rev1,rev2,rev3,rev4,rev5,rev6} including cold bosons
and their Bose-Einstein condensates as well as cold Fermi gases
like $^{40}$K and $^{6}$Li \cite{Wang,Cheuk}. Very recently, the
above fictitious spin-orbit coupling and Zeeman field in the
two-dimensional (2D) Fermi gas of $^{40}$K \cite{Huang} has been
reported. It has been shown that cold atom systems can generate
spin-orbit coupling and Zeeman-like splitting of the order of
the particles kinetic energy. To the best of our knowledge, such
high spin-orbit and Zeeman couplings can barely be realized in
condensed matter. If they do, this might open a venue to
qualitatively new manifestations of above interactions including
appearance of new quantum states of condensed matter.

One of the most interesting properties of cold matter is the
variety of interparticle interactions manifestations. For
Bose-Einstein condensates they are accurately described by the
Gross - Pitaevskii equation. For cold fermions, these
manifestations may be ferromagnetic fluctuations (the fermionic
isotope $^{6}$Li \cite{Jo}), at relatively strong interatomic
interactions. The effect of above fluctuations on the spin drag
was considered theoretically in While the actual quantum
phasemean field approximation in Ref. \cite{Duine}. Yet another
manifestation of the strong interactions between fermionic atoms
may be the realization of condensation - like phenomenon FC, see
Ref. \cite{book,pr,ks91}, generating flat (i.e. wave vector
${\bf k}$ independent so that corresponding quasiparticle cannot
propagate) portions in the quasiparticles (or real particles)
spectrum (energy-momentum relation). These flat portions
generate the deviation of the initial (i.e. that without FC)
Fermi distribution $n({\bf k})$ from step function
($n(k<k_F)=1$, $n(k>k_F)=0$, $k_F$ is Fermi wave vector) at
$T=0$. Namely, in these flat portions of the spectrum (i.e. in
FC phase) $0< n({\bf k}) <1$, while out of them $n({\bf k})$ is
either 1 or 0, see Refs. \cite{ks91,lidsky,yak,Yudin,book} for
details.  Recently, Yudin et al. \cite{Yudin} proposed to
observe FC - like effects (related to the emergence of flat
portions of the spectrum) in optical lattice systems of
ultracold fermions with a van Hove singularity in the Brillouin
zone.

While the actual quantum phase transition to the FC state is
envisioned to occur zero temperature, we demonstrate here a
two-dimensional atomic Fermi gases with the fictitious
spin-orbit coupling and magnetic field described above can
exhibit a finite-temperature state that may properly be regarded
as a precursor of a actual (or hypothetical) fermion
condensation at $T=0$. We remark that the entropy problem of
fermion condensation at $T=0$ is considered in Section
\ref{hts}, where we show that the entropy of the system in
question vanishes at $T\to0$ \cite{pr}. Our consideration of a
finite temperature state is dictated by the experimental
situation \cite{rev1,rev2,rev3,rev4,rev5,rev6}, in which very
low (1.5 $\cdot 10^{-7}$ K) but nonzero temperatures are
realized in the presence of unavoidable photon-atom collisions.
We proceed to construct the $H-\gamma$ phase diagram of the
system in which the FC precursor can potentially occur ($H$
being the fictitious Zeeman field and $\gamma$ the strength of
the spin-orbit interaction).  Also, we calculate the magnetic
moment and spin susceptibility as functions of the strength
$H_z$ of the artificial magnetic field.  The resulting
dependences have distinctive features in the range of magnetic
fields where FC precursor state can be realized.  Such features
may be considered experimental fingerprints of FC precursor
state. Another proposed fingerprint is of spectroscopic nature
and involves the photonic absorption spectrum.

\subsection{Theoretical formalism}
\subsubsection{Ideal 2D fermionic gas}
We begin with the Hamiltonian of ideal 2D gas of fermionic
particles (we use atomic units where $\hbar=1$)
\begin{equation}  \label{ham0}
\mathcal{H}_0=\frac{k_x^2+k_y^2}{2m}
+\alpha\left(k_x\sigma_y-k_y\sigma_x\right)+H_z\sigma_z.
\end{equation}
Here the first term is kinetic energy, second and third ones
are, respectively, the (artificial) spin-orbit and Zeeman
interaction terms. Accordingly, $\alpha$ is dimensional
(fictitious) spin-orbit interaction constant, $H$ is a (also
fictitious) magnetic field (in energy units) and
$\sigma_{x,y,z}$ are Pauli matrices.

The eigenvalues of the Hamiltonian \eqref{ham0} have the form
($k^2=k_x^2+k_y^2$)
\begin{equation}  \label{energ}
E_\pm=\frac{k^2}{2m}\pm\sqrt{H_z^2+\alpha^2k^2}.
\end{equation}

The corresponding normalized eigenvectors (spinors) read
\begin{eqnarray}
&&|->=\frac{1}{\sqrt{\alpha^2k^2+Q_{{\bf k}+}^2}}\left(%
\begin{array}{c}
\alpha k \\
-ie^{i\varphi_{\mathbf{k}}}Q_{{\bf k}+}%
\end{array}%
\right), \nonumber \\
&&|+>=\frac{1}{\sqrt{\alpha^2k^2+Q_{{\bf k}-}^2}}\left(%
\begin{array}{c}
\alpha k \\
-ie^{i\varphi_{\mathbf{k}}}Q_{{\bf k}-}
\end{array}
\right), \label{sp}
\end{eqnarray}
where $Q_{{\bf k} \pm}=H_z\pm \sqrt{H_z^2+\alpha^2k^2}$,
$\varphi_{\mathbf{k}}=\arctan(k_y/k_x)$.

\begin{figure}
\centering
\includegraphics[width=1.1\columnwidth]{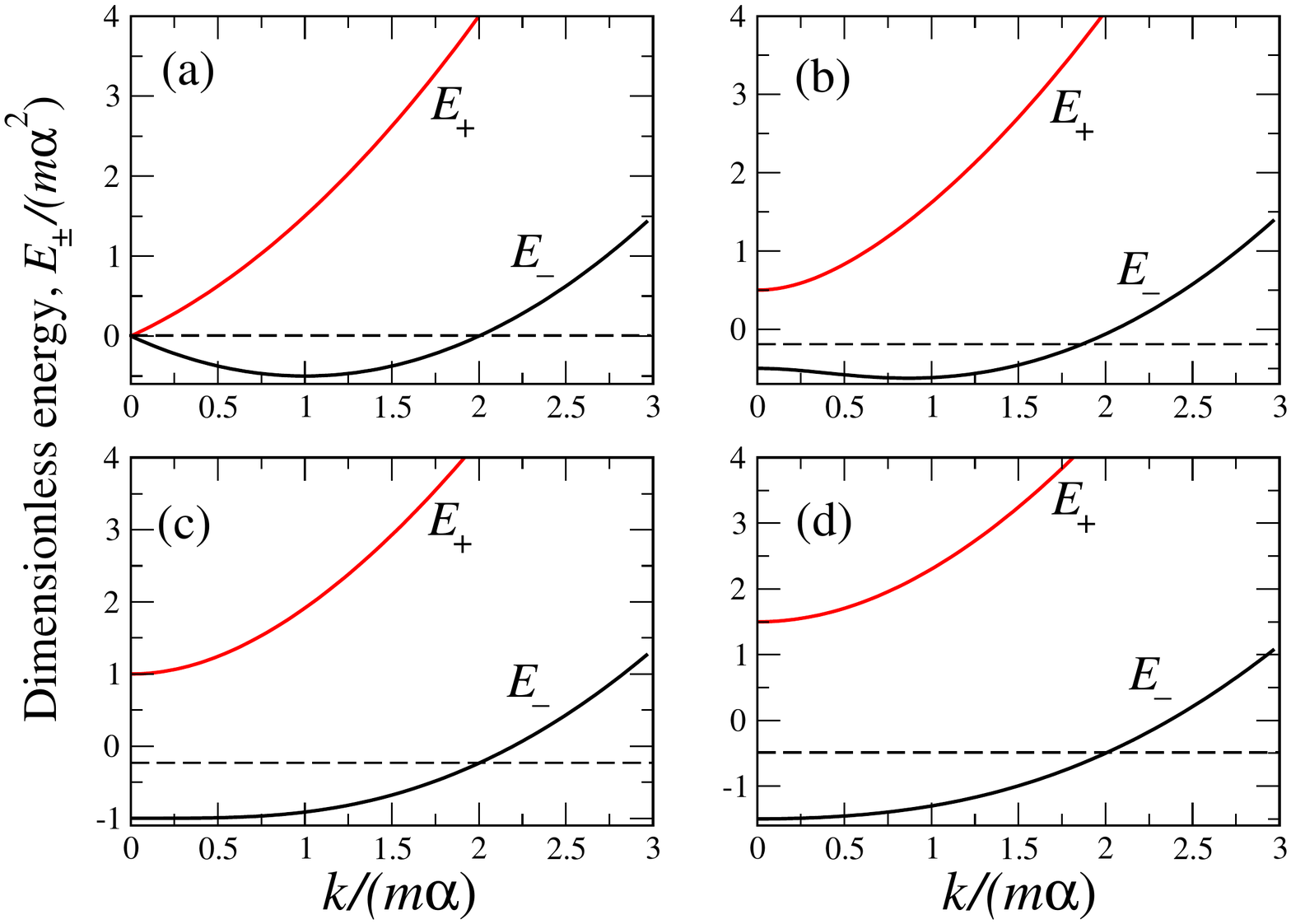}
\caption{(Color online) Spectrum of the ideal 2D gas of
fermionic atoms consisting of two branches $E_-$ (black curves)
and $E_+$ (red curves). The energies $E_\pm$ depend on the
artificial magnetic field strength $H_z$. Panels (a)-(d)
correspond to four cases:  (a) $H_z=0$, (b) $H_z=0.5m\alpha^2$,
(c) $H_z=m\alpha^2$ (critical value, when double-well character
of $E_-(k)$ disappears), and (d) $H_z=1.5m\alpha^2$. At $H_z
\neq 0$ a gap opens between $E_+$ and $E_-$. Horizontal dashed
lines correspond to the chemical potential $\mu$ at
$T=0.01m\alpha^2$, determined self-consistently from
Eq.~\eqref{fir} at the characteristic value $a=1/\pi$
corresponding to $\mu=0$ at $H_z=0$.  Occupied states are seen
to be present only for the $E_-$ branch.} \label{f1}
\end{figure}

It is instructive to study the spectrum of the noninteracting
gas \eqref{energ} as the classification of states remains the
same for a gas with interaction. Namely, the spectrum
\eqref{energ} has two branches, determined by $E_+(k,H_z)$) and
$E_-(k,H_z)$, where $E_+>E_-$ (Fig.\ref{f1}). It is seen from
Fig.\ref{f1} (see also Eq. \eqref{energ}) that at $H_z=0$ the
spectrum has minimum at $m\alpha$. At nonzero fields $H_z \neq
0$ a gap opens between $E_+$ and $E_-$, but at $H_z<m\alpha^2$
the lower branch $E_-$ still has shape with one maximum at $k=0$
and minimum. At high magnetic fields $H_z>m\alpha^2$ both
branches of the spectrum have a parabolic - like shape with one
minimum at $k=0$. Also, the chemical potential $\mu$ (see Eq.
\eqref{fir} below) is reported on the panels, showing that the
states of only the branch $E_-$ are occupied. This fact remains
valid also in the non-ideal, interacting fermionic gas case.

Simple analytical calculation confirms the above behavior.
Namely, the condition of extrema of $E_-(k,H_z)$ \eqref{energ}
yields two roots at $H_z \neq 0$
\begin{equation}
k_1=0, k_2=\sqrt{m^2\alpha^2-\frac{H_z^2}{\alpha^2}},
\label{kor1}
\end{equation}
where $k_1$ corresponds to a maximum, and $k_2$ to a minimum. At
$H_z=0$ the spectrum consists of branches of parabolas
$E_{\pm0}=k^2/(2m)\pm \alpha k$, with $E_{-0}$ having a minimum
at
 \begin{equation}\label{k0}
k_0= m\alpha.
 \end{equation}
The root $k_2$ \eqref{kor1} exists only if the expression under
square root is positive, i.e. $m^2\alpha^2>H_z^2/\alpha^2$ or
$H_z<H_0=m\alpha^2$. We note here that at $H_z=H_0$ the second
derivative $d^2E_{k-}/dk^2=0$, which means an infinite effective
mass in this point. We also calculate the values of $E_{k\pm}$
in the extrema to obtain
\begin{eqnarray}  \label{kor2}
E_{\mathbf{k}\pm}(k=0)=\pm H_z,\nonumber \\
E_{\mathbf{k}-}(k=k_2)=-\frac 12
\left[m\alpha^2+\frac{H_z^2}{m\alpha^2}\right],\ H_z< m\alpha^2,
\end{eqnarray}
which is in agreement with numerical calculations from Fig.
\ref{f1}.

To study further the occupation numbers $n_{\mathbf{k}\pm}$
\begin{equation}
n_{\mathbf{k}\pm}=\left\{1+\exp\left
[\frac{E_{\mathbf{k}\pm}-\mu}{T}\right]\right\}^{-1},
\label{nuf5a}
\end{equation}
we need to determine the chemical potential $\mu$, which is the
same for both $E_+$ and $E_-$ branches as our system is in
thermodynamic equilibrium. This can be done from the definition
of particles (or quasiparticles) density $\rho$ \cite{landau9}
\begin{equation}
\int
(n_{\mathbf{k}+}+n_{\mathbf{k}-})\frac{d^2k}{(2\pi)^2}=\rho,\
\rho=ak_0^2,\ 0<a<4.  \label{den1}
\end{equation}
Here we use the customary definition \cite{rev1} of cold
fermionic atoms density via wave vector $k_0$ \eqref{k0}. As in
the vast majority of experimentally realizable cases, the
occupation numbers even for interacting gas (see, e.g. Ref.
\cite{rev1}) depend on $k$ modulus only, the expression
\eqref{den1} can be further simplified after angular integration
to give
\begin{equation}
\int_0^\infty (n_{\mathbf{k}+}+n_{\mathbf{k}-})kdk=2\pi a\
m^2\alpha^2. \label{fir}
\end{equation}

The equation \eqref{fir} will be used now and subsequently (for
the interacting case) to determine the functions
$\mu_{\pm}(T,H_z)$ at finite temperatures.

\begin{figure}
\centering
\includegraphics[width=1.1\columnwidth]{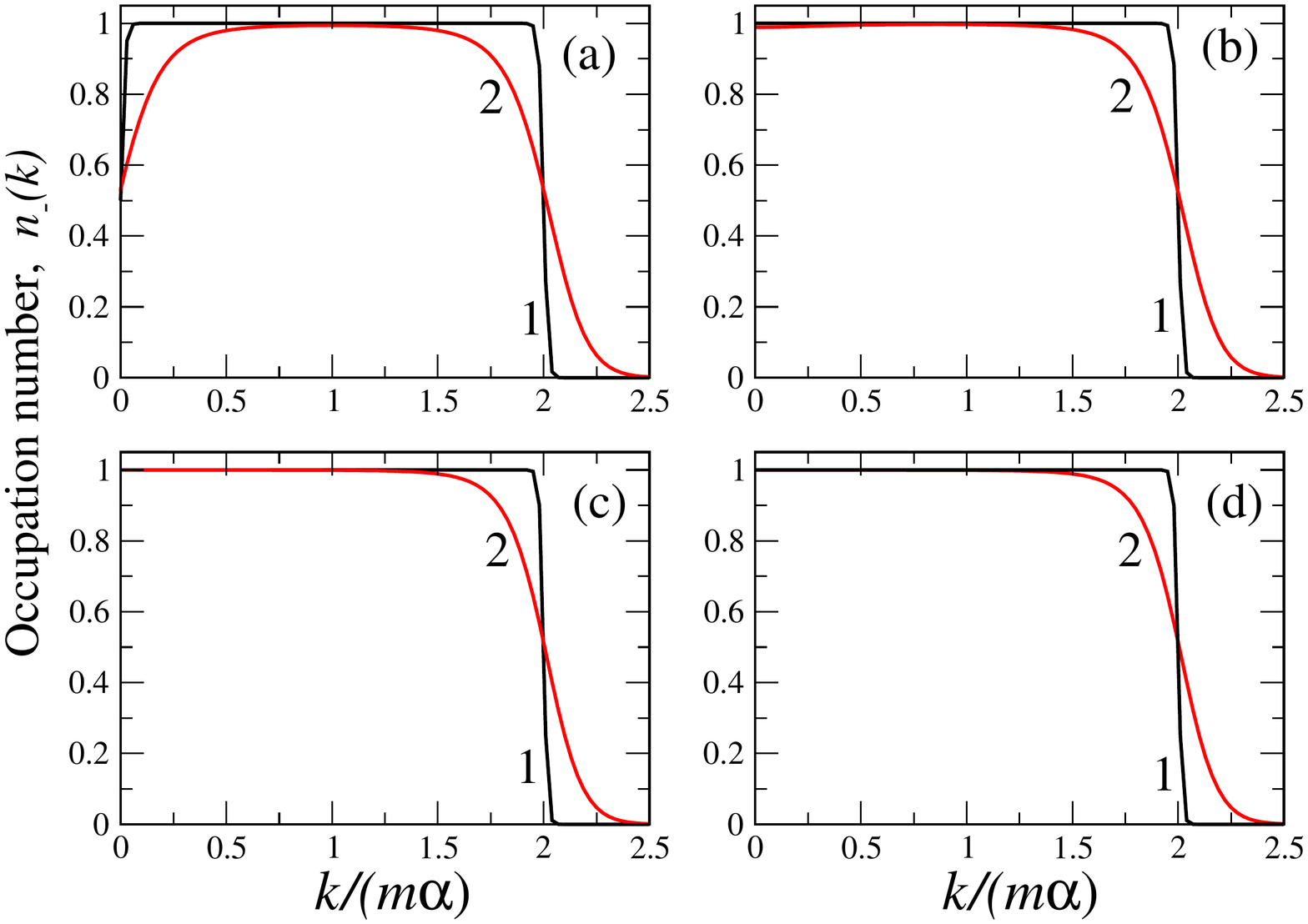}
\caption{(Color online) Occupation numbers $n_-(k)$,
corresponding to $E_-$ branch (the only having occupied states),
of the ideal 2D gas of fermionic atoms for two different
temperatures: $T_1=0.01m\alpha^2$ (black curves 1) and
$T_2=0.1m\alpha^2$ (red curves 2). Panels (a) - (d) correspond
to $H_z$ values similar to those from Fig. \ref{f1}. The
chemical potentials $\mu$ for $T=T_1$ are also the same as those
in Fig.1; $\mu(T_1)\approx \mu(T_2)$.} \label{f2}
\end{figure}

The occupation numbers $n_-(k)$, corresponding to the only
branch $E_-(k)$ with occupied states, are portrayed in
Fig.\ref{f2}. It is seen, that (fictitious) spin-orbit
interaction generates non-typical dependence $n_-(k)$. Really,
while normally $n(k)$ has step ($T=0$) or "blurred step" ($T
\neq 0$) behavior, at $H_z<m\alpha^2$ (when the spin-orbit
interaction is important, which is most pronounced at $H_z=0$)
we have $n_-(k)$ to be of "$\Pi$" shape.  At high magnetic
fields $H_z>m\alpha^2$, the spin-orbit interaction becomes
negligible and the ordinary shape of $n_-(k)$ restores. The
latter features of $n_-(k)$ shape is also inherent in the
interacting case. Below we will also see that FC precursor
behavior with flat, dispersionless portions of the spectrum
$E_-(k)$ (our analysis shows that for the problem under
consideration, the branch $E_+(k)$, even if occupied, never has
FC-like behavior) will occur both at $H_z<m\alpha^2$  and
$H_z>m\alpha^2$ depending on constant $a$ (responsible for atoms
concentration in a gas) in the expression \eqref{den1}.

\subsubsection{Interacting fermionic gas}

The ground state energy of the interacting, non-ideal gas of
cold fermionic atoms can now conveniently be written in the
basis \eqref{sp}, diagonalizing the ideal fermionic gas
Hamiltonian \eqref{ham0}. In the spirit of Ref. \cite{rev1}, it
consists of the ideal gas spectrum \eqref{energ} and all
possible matrix elements $g_{{\bf k}, {\bf k'},\alpha,\beta}$
$\equiv <\alpha,{\bf k}'|\beta, {\bf k}>$ ($\alpha, \beta=\pm$),
playing a role of interaction term. To be specific, the ground
state energy functional reads
\begin{eqnarray}
W &\equiv& <\mathcal{H}-\mu N>_G =
\sum_{\mathbf{k}}\Bigl[(E_{\mathbf{k}-}-\mu)n_{\mathbf{k}-}
+(E_{\mathbf{k}+}-\mu)n_{\mathbf{k}+}\Bigr]+\nonumber \\
&&+\gamma \sum_{\mathbf{k},\mathbf{k}^{\prime }}
\Bigl[g_{\mathbf{k}\mathbf{k^{\prime }}++}\ n_{\mathbf{k}+}
n_{\mathbf{k^{\prime }}+}+g_{\mathbf{k}\mathbf{k^{\prime }}+-}
\ n_{\mathbf{k}+} n_{\mathbf{k^{\prime }}-} \nonumber \\
&&+g_{\mathbf{k}\mathbf{k^{\prime }}-+}\ n_{\mathbf{k}-}
n_{\mathbf{k^{\prime }}+} +g_{\mathbf{k} \mathbf{k^{\prime
}}--}\ n_{\mathbf{k}-} n_{ \mathbf{k^{\prime }}-}\Bigr].
\label{nuf2}
\end{eqnarray}
Here $\gamma$ is the interaction constant carrying the
dimensions of energy and $E_{{\bf k},\pm}$ is determined by Eq.
\eqref{energ}. Note that while at finite temperatures we should
minimize the functional $U=W-TS$, where $T$ is a temperature and
$S$ is an entropy (see Ref. \cite{book} and references therein
for details), for the experimentally realizable temperatures
$T_0 \sim 1.5 \cdot 10^{-7}$ K it is sufficient to minimize the
ground state energy only, however, with $n_{\mathbf{k}\pm}$
being temperature dependent by virtue of Eq. \eqref{nuf5a}.

The matrix elements $g_{{\bf k}, {\bf k'},\alpha,\beta}$ may be
expressed as
\begin{eqnarray}
&&g_{\alpha, \beta}(x,x')=\frac{xx'}{S_\alpha(x) S_\beta(x')},\nonumber \\
&&S_\pm(x)=\sqrt{x^2+(h \mp \sqrt{h^2+x^2})^2}. \label{mel}
\end{eqnarray}
having introduced dimensionless variables $x=k/(m\alpha)$ and
$h=H_z/(m\alpha^2)$. We note that the expressions for matrix
elements \eqref{mel} contain an additional term proportional to
$\cos \varphi_{\bf k}$. For isotropic solutions (i.e. dependent
only on modulus $k$), this part gives zero after integration
over $\varphi_{\bf k}$.

Going from summation to integration in Eq.~\eqref{nuf2} and
varying over $n_{{\bf k}\pm}$, we arrive at the following set of
integral equations for determination of resulting spectrum and
occupation numbers
\begin{eqnarray}
E_{\mathbf{k}+}-\mu &+&\frac{\gamma}{(2\pi)^2}\int_0^\infty
k'dk'
\int_0^{2\pi}d\varphi_{\bf k'} \times \nonumber \\
&&\Bigl[g_{{\bf k} {\bf k'}++}\ n_{{\bf k'}+} + g_{{\bf k}  {\bf
k'}+-}\
n_{{\bf k'}-}\Bigr]=0,  \nonumber \\
E_{\mathbf{k}-}-\mu &+&\frac{\gamma}{(2\pi)^2}\int_0^\infty
k'dk'
\int_0^{2\pi}d\varphi_{\bf k'} \times \nonumber \\
&&\Bigl[g_{\mathbf{k}
\mathbf{k^{\prime }}-+}\ n_{\mathbf{k^{\prime }}+}+g_{\mathbf{k}\mathbf{%
k^{\prime }}--} \ n_{\mathbf{k^{\prime }}-}\Bigr]=0.
\label{nuf4}
\end{eqnarray}
Here the occupation numbers $n_\pm$ are related to the spectrum
$E_\pm$ by the expression \eqref{nuf5a}. The system \eqref{nuf4}
should be augmented by the equation for self-consistent
determination of chemical potential $\mu\equiv $  $\mu(H_z,T)$
\eqref{fir}.

\subsection{Results and discussion}

To make our (numerical) solution of the set \eqref{nuf4} to be
physically meaningful, we estimate some characteristic
parameters of 2D gas of cold fermionic atoms on the base of
experimental situation \cite{rev1,rev2,rev3,rev4,rev5,rev6}.
Typical spin-orbit coupling constant in the above gases is
$\alpha =10$ cm/s. In our calculations we will focus primarily
on the fermionic isotope $^6$Li with mass $m=10^{-23}$ g. For
this atom the characteristic energy related to spin - orbit
interaction is $E_c=m\alpha ^2=10^{-21}$ erg or $6\cdot 10^{-6}$
meV, which corresponds to temperature $\sim 10^{-5}$ K. The
related momentum $k_c$ and concentration $\rho_c$ units are,
respectively $k_c=m\alpha =10^5$ cm$^{-1}$ and $\rho_c =a
k_c^2\approx a\cdot 10^{10}$ cm$^{-2}$, where parameter $a$ is
defined in Eq. \eqref{den1}.

The interaction constant $\gamma$ can be estimated as
$\gamma=4\pi (a_s/w)E_c=4\pi (a_s/w)M\alpha^2$ \cite{rev1}. Here
$a_s$ is a three-dimensional scattering length, which is of the
order of 100 Bohr radii. For two-dimensional case it should be
divided by the width of the optically trapped layer ("pancake")
$w$ which is typically several microns. With this in mind, the
characteristic value of the interaction constant is $\gamma=4\pi
\cdot 0.0053E_c=7\cdot 10^{-23}$ erg $=5\cdot 10^{-7}$  Kelvins.
This value is by order of magnitude the same as the
characteristic experimentally realizable temperature $T=1.5\cdot
10^{-7}$ K due to photon-atom energy transfer. These numerical
estimations show that physically meaningful values of the
dimensionless temperature are $\tau=T/(M\alpha^2) \sim 0.01 -
0.05$, while other parameters like dimensionless magnetic field
$h$ and interaction constant $\gamma_0=\gamma/(M\alpha^2)$ could
be varied to achieve the desired effect.

\begin{figure}
\centering
\includegraphics[width=1.1\columnwidth]{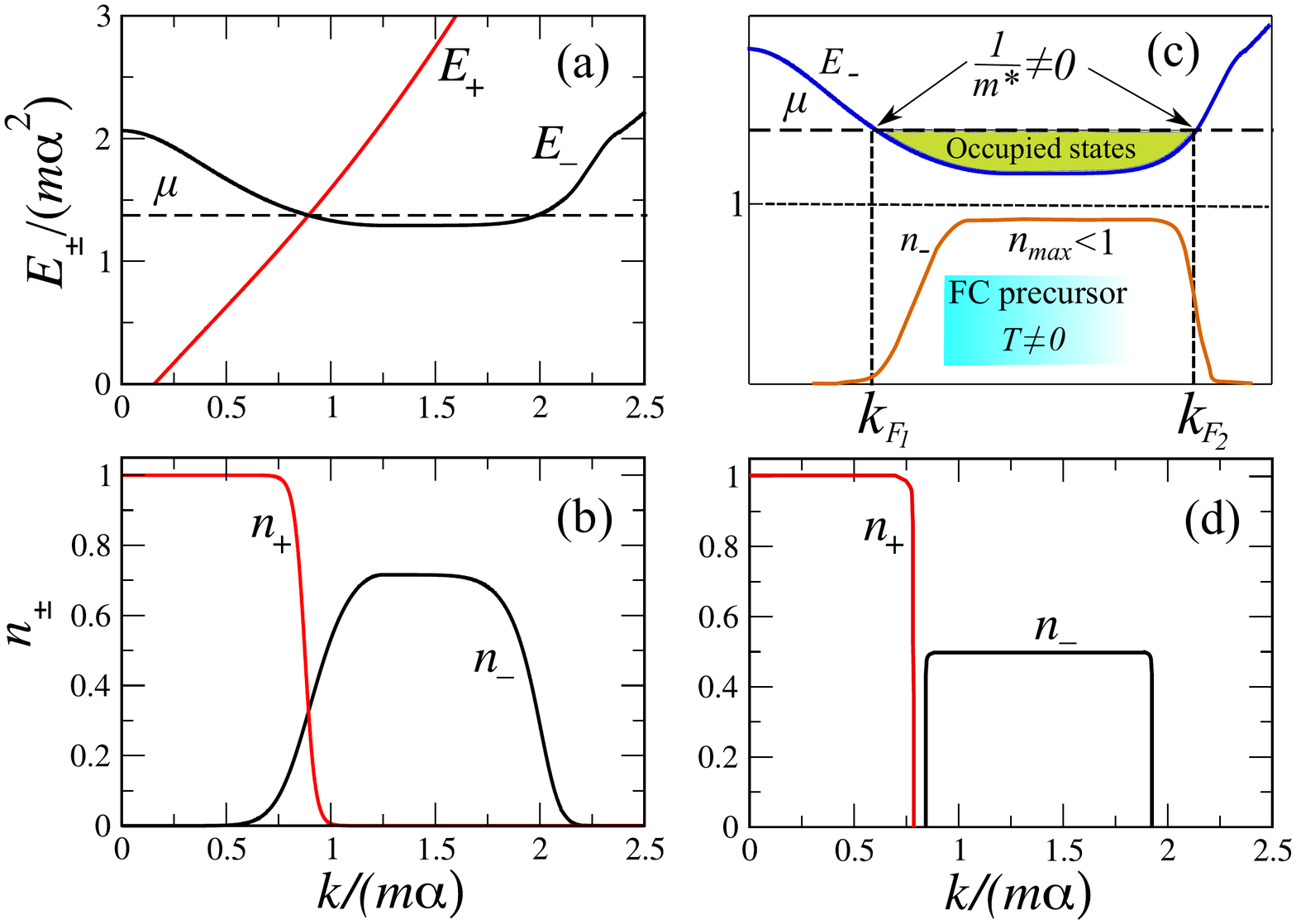}
\caption{(Color online) The numerical solution of the set
\eqref{nuf4} corresponding to the FC precursor realization. The
spectrum $E_\pm (k)$ is reported in the panel (a) and
corresponding occupation numbers $n_\pm (k)$ in panel (b). The
solution is obtained for $H_z=0.5m\alpha^2$, $T=0.05m\alpha^2$
and $\gamma=12m\alpha^2$. This correspond to the chemical
potential $\mu=1.33785m\alpha^2$ and bottom (flat part) of the
$E_-(k)$ $E_{-min}=1.2917m\alpha^2$. It is seen that FC
precursor occurs for $E_-(k)$, while $E_+(k)$ exhibits standard
behavior.  {Panel (c) shows schematically energy spectrum and
occupation numbers for the FC precursor state realized at $T
\neq 0$. Panel (d) reports the FC precursor state at $T=0$, see
text for details.}} \label{f3}
\end{figure}

The set of integral equations \eqref{nuf4} has been solved
iteratively with respect to Eq. \eqref{fir} in the above
dimensionless variables. The natural choice of zeroth approach
in this case is ideal gas spectrum \eqref{energ}. The results
are shown in the Fig. \ref{f3} in the form of $E_\pm (k)$ (a)
and $n_\pm (k)$ (b) dependences.
 {We choose the parameters $H_z$
and $\gamma$, where FC precursor is realized. The temperature
(except Fig. \ref{f3}d, where $T=0$) has been chosen to be
$T=0.05m\alpha^2$ as this value reflects the experimental
situation for $^6$Li atom.} It is seen, that latter phenomenon
occurs for the branch $E_-(k)$ only, while $E_+(k)$ remains
intact. We did not find contradictions for this regularity: for
all possible $H_z$ and $\gamma$, the FC (if any) occurs for
$E_-(k)$ only. This shows the importance of spin - orbit
interaction in the process of FC precursor formation.

The main peculiar feature of our FC precursor phenomenon is the
emergence of the flat, dispersionless portion $E_{-min}$ in the
bottom of $E_-(k)$ branch with chemical potential $\mu$ being
very close to it. This implies (Fig. \ref{f3}b) that
corresponding occupation number $n_-(k)$ has its maximal value
$n_{max}$ to be less then 1.  {If we put $T \to 0$ (which is not
the case for all possible $H_z$ and $\gamma$ where FC precursor
occurs), the corresponding dependence $n_- (k)$ will be of
exactly "$\Pi$" shape with $n_{max}=0.5$, Fig. \ref{f3}d. Panel
(c) of Fig. \ref{f3} reports the schematics of energy spectrum
and occupation number $n_-(k)$ for our precursor FC state at low
but nonzero temperatures.} The main feature realized in the FC
quantum phase transition at $T=0$ is exact equality $E(k)=\mu$
at some portion of the spectrum $E(k)$ limited by two wave
vectors $k_i$ and $k_f$ (index "i" stands for "initial" and "f"
for final),  {see the book \cite{book} for pedagogical
introduction.} This, in turn, implies the nonanalytic behavior
of $E(k)$ at $k_i<k<k_f$ with all derivatives $d^nE(k)/dk^n=0$.
The zeroth second derivative $d^2E(k)/dk^2=1/m^*=0$ means the
infinite fermion effective mass $m^* \to \infty$, which is
peculiar experimentally observable feature of FC in solids
\cite{book,ks91}. The occupation number for FC also behaves
differently from both standard Landau Fermi liquid
\cite{landau9} and our FC precursor phase. Namely, its maximal
value $n_{max}=1$ with subsequent gradual decay down to $n=0$.
The decay region is exactly $k_i<k<k_f$ so that initial Fermi
wave vector $k_F$ is hidden (or disappears) in FC state. It has
been shown by Volovik \cite{vol1,vol2,vol3}, that this is the
consequence of the Fermi surface topology altering in the Fermi
condensation point at T=0.  {This topology altering can be
understood as follows. While the region of the occupied states
(between flat part of the spectrum and chemical potential $\mu$)
shown in Fig. \ref{f3}c has finite width, the corresponding
region in real FC state \cite{book} is of infinitesimal width.
In other words, while our FC precursor can be viewed as "shallow
Fermi sea" (between $k_{F_1}$ and $k_{F_2}$ on Fig.\ref{f3}c),
the real FC phase can be regarded as a "beach of Fermi sea". We
note that the qualitative FC features are the same for any
potential of inter-fermion interaction, yielding real FC
phenomenon \cite{book,ks91}. Also, the analysis of "FC traces"
at finite temperatures \cite{book} shows that the system begins
to enter into FC phase at lowering temperatures $T \leq T_f$
when $T \sim E_{min}-\mu$, with $T_f$ being the temperature at
which FC starts to define the properties of system in question
\cite{book,pr}.} As for calculation from Fig.\ref{f3}a, b
$E_{min}/(m\alpha^2)=1.2917$ and $\mu/(m\alpha^2)=1.33785$,
their difference 0.046 is approximately the temperature
$T/(m\alpha^2)=0.05$. This permits us to hope that our FC
precursor has to do with real FC phenomenon at $T=0$.

 {The fact that at low
temperatures the spectrum $E_{\mathbf{k}-}-\mu \sim T$ can also
be shown using following analytical arguments. As both types of
occupation numbers $n_{\mathbf{k}\pm}$ obey Fermi-Dirac
statistics \eqref{nuf5a}, the corresponding spectra can be
represented as
\begin{equation}\label{nw1}
E_{\mathbf{k}\pm}-\mu(T)=T\ln\frac{1-n_{\mathbf{k}\pm}}{n_{\mathbf{k}\pm}}.
\end{equation}
At low temperatures, say $T<0.05m\alpha^2$, the occupation
numbers $n_{\mathbf{k}\pm}$ can be approximately considered as
rectangles, see Fig.\ref{f3}b, d. This permits to approximately
evaluate the integrals in Eqs \eqref{nuf4} to constants $C_\pm$,
where signs correspond to the equations with $E_{\mathbf{k}\pm}$
respectively. Our numerical calculations show that at
$T<0.05m\alpha^2$ constants $C_\pm$ are weakly temperature
dependent. Also, as $T \to 0$ the constant $C_-$ goes to zero
while $C_+$ remains almost the same. This implies that at $T \to
0$ $E_{\mathbf{k}-}-\mu(T)\approx
T\ln((1-n_{\mathbf{k}-})/n_{\mathbf{k}-}) \sim T$ as for low
temperatures $n_{\mathbf{k}-}$ has almost $\Pi$ shape (Fig.
\ref{f3}b and d) and can be considered as a constant at some
interval of $k$.}

At the same time, although in our FC precursor phase the
spectrum $E_-(k)$ also has nonanalytic behavior with flat part,
this does not imply the infinite effective mass. Really, as it
is seen from Fig.\ref{f3}c, this nonanalytic region does not
fall into the intersection between $E_-(k)$ and $\mu$. At these
intersection points, labeled $k_{F_1}$ and $k_{F_2}$, the slope
of $E_-(k)$ is finite and hence the effective mass is finite
also. We consider this finiteness of the effective mass to be
the main difference between our FC precursor and real FC in
solids. We call our phenomenon "precursor" as at $T \to 0$ and
eventual $E_{-min} \to \mu$ many features of real FC will be
 {realized, see Fig.\ref{f3}d. One
of the differences is "$\Pi$" shape of the occupation numbers
$n_-(k)$ (Fig.\ref{f3}d),} which is the consequence of
spin-orbit interaction presence. Note that the behavior of
occupation numbers both in our FC precursor and in real FC
phases does not contradict the Pauli exclusion principle as in
both cases $0<n<1$.

\begin{figure}
\centering
\includegraphics[width=0.9\columnwidth]{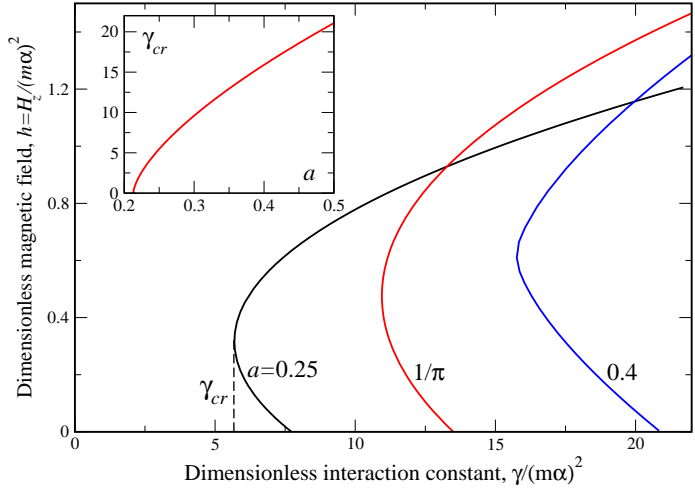}
\caption{(Color online) The phase diagram of the system under
consideration in the variables $h-\gamma$ at $T=0.05m\alpha^2$
and different concentration parameters $a$ (Eq. \eqref{den1},
numbers near curves). The FC precursor is realized inside (i.e.
between the branches) of parabola-like curves at each $a$.
Inset portrays the dependence of critical interaction constant
$\gamma_{cr}$ (taken at the curves vertices, shown as an example
for the curve $a=0.25$ by dashed line) on parameter $a$. At
$a<0.213$ the FC precursor does not exist at any $h$ and
$\gamma$. (Color figure online.)} \label{f4}
\end{figure}

The next natural question is at which $H_z$ and $\gamma$ (at a
given $T$) the above FC precursor phenomenon occurs. The phase
diagram of the system under consideration in the dimensionless
variables $h=H_z/(m\alpha^2)$ - $\gamma/(m\alpha^2)$ is reported
in Fig. \ref{f4}. The FC precursor phase exists between the
branches of parabola-like curves $h(\gamma)$, corresponding to
certain concentration parameter $a$. For instance, at $a=1/\pi$
(corresponding to $\mu=0$ for $\gamma=0$ and $H_z=0$) and
$\gamma=12$, the FC precursor exists in the field interval
$0.16<h<0.79$, see Fig.\ref{f4}. It is seen that for large
$\gamma$, corresponding to strong interatomic interaction, the
FC precursor state starts to exist already at $H_z=0$. On the
other hand, large fields suppress the latter state, which agrees
with the behavior of real FC state in solids \cite{book}.
Actually, the property that FC precursor exists in our system in
some magnetic field interval $H_{min}<H_z<H_{max}$ is in accord
with the experimentally observed properties of solids, which can
be explained in terms of fermion condensation, see, e.g. Fig.7
of Ref. \cite{annphys}. Note that at strong interactions the FC
precursor state exist also at $H_z >m\alpha^2$, when the
dispersion $E_-(k)$ minimum shifts to $k=0$ and spin-orbit
interaction becomes unimportant.

The vertex of parabola-like curves in Fig.\ref{f4} determines
the critical interaction $\gamma_{cr}$ such that at a given $a$
and $\gamma<\gamma_{cr}$ the FC precursor does not exist. This
is because the weak interaction cannot organize the system into
collective FC precursor state. The dependence $\gamma_{cr}$ on
concentration parameter $a$ is reported in the inset to
Fig.\ref{f4}. The existence of threshold concentration parameter
$a_0=0.213$ is seen. This shows that for FC precursor
realization in the system under consideration the concentration
of atoms should be more then threshold value, determined by the
parameter $a_0$. In other words, to gain such collective state,
as FC precursor, we need to have sufficient concentration of
strongly enough interacting fermionic atoms. The strong
(fictitious) magnetic fields in this case only suppress this
collective state similarly to the real magnetic field in solids
\cite{book}. Our analysis shows that the dependence
$\gamma_{cr}(a)$ approaches $a_0$ with infinite derivative and
that the entire phase diagram depends on temperature. Latter
dependence is weak at $T/(m\alpha^2)<0.05$ so that qualitative
features of the phase diagram remain the same at $T \to 0$.

\begin{figure}
\centering
\includegraphics[width=1.0\columnwidth]{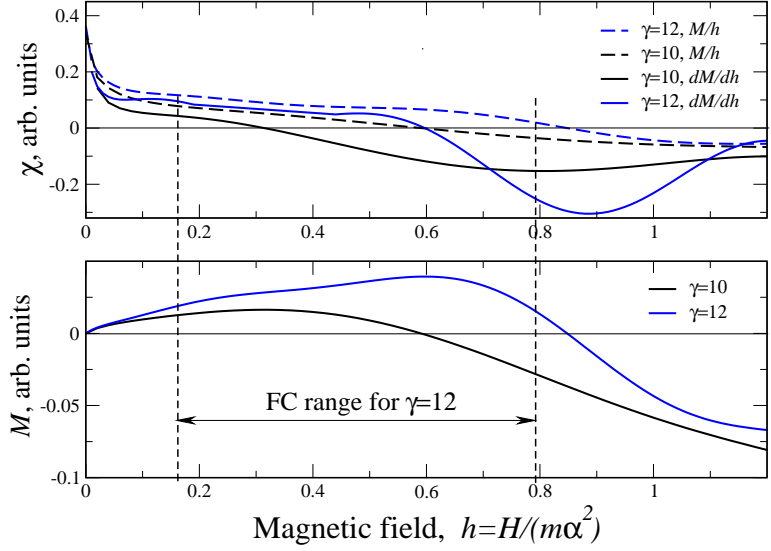}
\caption{(Color online) The spin susceptibilities (upper panel)
and magnetization (lower panel) of the system under
consideration as functions of dimensionless magnetic field $h$
at $T=0.05m\alpha^2$. Black curves - $\gamma=10$ (no FC
precursor), blue curves - $\gamma=12$ (the range of magnetic
fields, where FC precursor occurs, is shown). Dashed lines
correspond to $M/H$ and full lines - $dM/dH$. (Color figure
online.)} \label{f5}
\end{figure}

The frequently experimentally observed quantity in the cold
fermionic gases is their spin magnetization
\begin{eqnarray}
M=S_z =\frac 12\int \Bigl[ n_{+,k}\sigma_{z++}+n_{-,k}
\sigma_{z--}\Bigr] \frac{d^{2}k}{(2\pi)^{2}} =\nonumber \\
=\frac{H_z}{2}\int_0^{\infty }\frac{n_{+,k}-n_{-,k}}
{\sqrt{H_z^2+\alpha ^2k^2}}\ \frac{kdk}{2\pi}. \label{magn}
\end{eqnarray}
Here $\sigma_{z++}\equiv \langle+|\sigma_z|+\rangle$ and
$\sigma_{z--}\equiv \langle-|\sigma_z|-\rangle$, where
$\sigma_z$ is Pauli matrix and $|+>$ and $|->$ are the states
\eqref{sp} of ideal fermionic gas Hamiltonian. The results of
calculations of $S_z$ as well as spin susceptibilities
$\chi=M/H_z$ and $\chi_{dif}=dM/dH_z$ at $T/(m\alpha^2)=0.05$
are reported in Fig. \ref{f5}. It is seen that qualitative
behavior of magnetizations and susceptibilities is almost
similar for the case of presence ($\gamma=12$) or absence
($\gamma=10$) of the FC precursor behavior. The main difference
is the "hump" in the magnetization near the upper magnetic field
boundary of FC precursor existence at $\gamma=12$. This "hump"
and subsequent faster (then that at $\gamma=10$) decay of
magnetization in this region generates a "well" in the
differential susceptibility $dM/dh$, see upper panel of
Fig.\ref{f5}. The magnetization "hump" and corresponding
differential susceptibility "well" near the upper field boundary
of FC precursor existence can be well regarded as possible
experimental "fingerprints" of this behavior. Although the same
maximum and minimum take place also for "no FC" case
$\gamma=10$, they are much less pronounced then those for
$\gamma=12$ and situated at much smaller fields. Our
calculations at $T/(m\alpha^2)=0.025$ show that both
magnetization and susceptibilities are almost the same as those
at $T/(m\alpha^2)=0.05$. This confirms our statement that above
peculiar features survive at temperature lowering and thus may
be regarded as experimental manifestations of possible FC
precursor realization in ultracold fermionic atom gas. To
determine the fields range, when possible FC precursor is
realized experimentally, the detailed comparison between
experimental and theoretical magnetization and susceptibility
curves are needed.

The important spectroscopic manifestation of the FC precursor
state is the study of the photonic absorption spectra, which
intensity $I$ can be related to the occupation numbers $n_+$ and
$n_-$ in a simple manner

\begin{eqnarray}
I(\omega)=\frac 12\left[n_+(1-n_-)+n_-(1-n_+)\right],\label{spp} \\
\omega \equiv \omega(k)=|E_-(k) - E_+(k)|.\nonumber
\end{eqnarray}
The above absorption spectrum defines the transitions (at a
given wave vector and hence the frequency \eqref{spp}) from
occupied states in either $E_+$ or $E_-$ branch of the spectrum
to the corresponding free states at a given $k$. As the
frequency of such transition should always be positive, it is
defined as a modulus. To understand better, how these
transitions occur, we take a look on Fig.\ref{f3}a. At small $k$
the states $E_+<\mu$ are occupied so that transitions go from
$E_+$ to $E_-$. After the point $E_+=\mu$, which is almost the
same as $k_{F_1}$ in Fig.\ref{f3}c we have the opposite
situation, where the transitions occur from occupied $E_-$
states to free $E_+$ ones. Such situation is realized until $k$
reaches $k_{F_2}$ (Fig.\ref{f3}c), whereupon both $E_+$ and
$E_-$ states become free so that the transitions are impossible.
This means that point $k_{F_2}$, where the lower energy branch
equals $\mu$ is the absorption spectrum termination point.

\begin{figure}
\centering
\includegraphics[width=1.0\columnwidth]{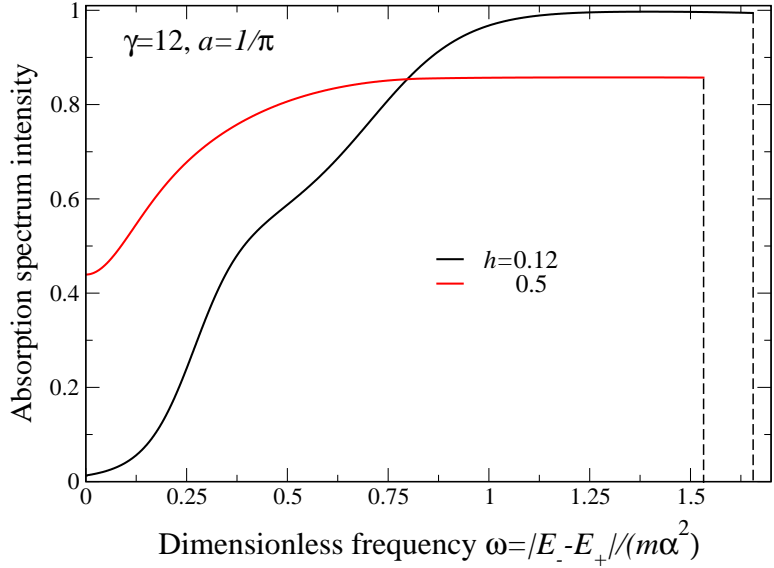}
\caption{(Color online) The photonic absorption spectra of the
system under consideration at $T=0.05m\alpha^2$, $a=1/\pi$ and
$\gamma=12$. Black curves - $h=0.12$ (no FC precursor), red
curves - $h=0.5$. Vertical dashed lines correspond to spectrum
termination points: 1.919 for $h=0.12$ and 1.678 for $h=0.5$.
(Color figure online.)} \label{f6}
\end{figure}

With this in mind, in Fig. \ref{f6} we plot the absorption
spectrum for two magnetic field values: $h=0.12$ (no FC
precursor) and $h=0.5$. The rest of parameters are $a=1/\pi$,
$\gamma=12$ and $T/(m\alpha^2)=0.05$. The main difference
between FC precursor absence and presence cases is that while in
former case the saturation value of absorption line equals one
(hence the factor 1/2 in the definition of $I(\omega)$
\eqref{spp}), in the latter case this value is always less then
one. The origin of such behavior can be seen in Fig.\ref{f3}b,
where $n_-$ for FC precursor phase is plotted. Namely, in this
case the saturation value $n_{max}<1$, see also Fig.\ref{f3}c.
Substitution of this inequality into the expression \eqref{sp}
even for $n_+=1$ immediately generates $I_{max}<1$. This means
that most pronounced experimental manifestation of possible FC
precursor in ultracold fermionic gases is the fact that
saturation amplitude of photonic absorption spectrum is less
then unity. The spectra terminate at respective values of
$k_{F_2}$ in dimensionless frequency units: 1.919 for $h=0.12$
and 1.678 for $h=0.5$.

\subsection{Conclusions}
In this second major section, we have devised and explored a
simple but realistic model of a 2D gas of cold fermionic atoms,
which, at experimentally attainable low temperatures, is found
to exhibit features characteristic of a precursor of fermion
condensation. We have shown that the presence of a spin-orbit
interaction is necessary for the emergence of this new
collective state.  We have addressed the similarities and
differences of and between this state and the real, or fully
developed, fermion-condensation phenomenon, which has been
studied since the early 1990's \cite{ks91,book}. The potential
experimental manifestations of possible FC precursor formation
have also been considered at some depth.  In particular, we
predict that in a FC precursor phase the spin magnetization and
susceptibility possess specific signatures as functions of the
fictitious magnetic field, in the form of a deep minimum of the
differential susceptibility.  Another predicted manifestation
concerns photon absorption spectra, in which case the saturation
value near the termination point should become less than unity.
To support these predictions, we have estimated typical system
parameters required for realization of the new collective state
associated with such FC precursor behavior in ultracold
fermionic gases, with the distinct prospect of future
experimental confirmation, as the required conditions are
expected to be realistically accessible.

\section{Overdoped high-$T_c$ superconductors}\label{hts}

\subsection{Introduction}

Strongly correlated Fermi systems hosting a fermion condensate
(FC) exhibit unusual properties in both superconducting and
(putatively) normal phases. From the fundamental perspective,
the presence of a FC breaks both particle-hole symmetry and the
time-reversal invariance\cite{jetp2003,shagstep,tun,baras}. As a
profound consequence, the well-know Leggett theorem \cite{leg}
is violated.  This theorem states that in any superconducting
state of an electronic fluid at $T=0$, the number density of
superconducting electrons, $n_s$, is {\it equal} to the total
electron number density $n_e$.

Recent experimental studies of overdoped high-$T_c$
superconductors (HTSC) have revealed strong deviations of their
physical properties from those predicted by canonical Bardeen-
Cooper-Schrieffer (BCS) theory \cite{bosovic}.  Especially
confounding, from the conventional vantage, is the documented
failure of Leggett's theorem \cite{zaanen}. The observed
deviations were found to be surprisingly similar in numerous
HTSC samples \cite{bosovic,zaanen,uemura,uem_n,bern,rour}.
Measurements of the absolute values of the magnetic penetration
depth $\lambda$ and the phase stiffness $\rho_s=A/\lambda^2$
were carried out on thousands of virtually perfect
two-dimensional (2D) samples of $\rm La_{2-x}Sr_xCuO_4$ (LSCO)
under variation of doping $x$ and temperature $T$.  (Here
$A=d/4k_Be^2$, where $d$ is the film thickness, $k_B$ Boltzmann
constant, and $e$ the electron charge.) It was observed that the
dependence of the zero-temperature superfluid density (the
density of superconductive electrons), written as $n_s=4\rho_s
k_Bm^*$ where $m^*$ is the electron effective mass, is
proportional to the critical temperature $T_c$ over a wide
doping range. This dependence agrees with pervious measurements,
but is incompatible with the standard BCS description. Most
significantly, $n_s$ turns out to be considerably smaller than
the BCS density $n_{el}$ of superconductive electrons
\cite{bosovic,zaanen,uemura,uem_n,bern,rour}, which is
approximately equal to the total electron density
\cite{bardeen}. These observations, clarifying intrinsic
properties of LSCO, have provided unique opportunities for
testing and expanding our understanding of the physical
mechanisms responsible for high-$T_c$ superconductivity.  Our
intention here is to show that the physical mechanism
responsible for such clear departures from BCS behavior in
overdoped LSCO, stems from the topological phase transition
giving rise to the FC phenomenon that generates flat bands
\cite{book,ks91,vol1,vol3,volovik:2015,khodel:1994,pr}. We
propose that flat bands and an associated extended saddle-point
singularity play an important role in the theory of HTSC, as
substantiated in
Refs.~\cite{volovik:2015,khodel:1994,pr,qp2,abrikosov}.

Employing a formalism that accounts for the pertinent
fermion-condensation quantum phase transition (FCQPT), we shall
now investigate the overdoped LSCO system and show that as soon
as the doping $x$ reaches the critical value $x_c$ of this
transition, the features of emergent superconductivity begin to
differ from those of BCS theory. We will demonstrate that: (i)
at $T=0$, the superfluid density $n_s$ turns out to be only a
small fraction of the total density of electrons; (ii) the
critical temperature $T_c$ is controlled by $n_s$ rather than by
doping, being a linear function of $n_{s}$. Since the FCQPT
generates flat electronic bands
\cite{book,ks91,vol1,vol3,volovik:2015}, the system under
consideration exhibits non-Fermi liquid (NFL) behavior, notably
a resistivity $\rho(T)\propto \alpha T$ varying linearly with
temperature.   With $x\to x_c$, the factor $\alpha$ diminishes
with decreasing $T_c$, and the system then exhibits Landau
Fermi-liquid (LFL) behavior at $x>x_c$ and low temperatures.
These predictions are in good agreement with the recent
experimental findings \cite{bosovic,zaanen,pagl}.

\subsection{Two-component system}

Condensed matter theorists been have facing something of a
dilemma in attempting to explain the NFL behavior observed in
HTSC beyond the critical point where the low-temperature density
of states $N(T\to 0)$ diverges and flat bands can be generated
without breaking any ground-state symmetry.  (For relevant
background and developments, see Refs.~
\cite{bosovic,volovik:2015,khodel:1994,pr,pagl,khod:2015,lifshitz}.)
In homogeneous matter, such a divergence is associated with the
onset of a topological transition at $x=x_c$ signaled by the
appearance of an inflection point at momentum $p=p_F$
\cite{khodel:1994,pr,ybalb}
\begin{eqnarray}\label{top}
\varepsilon-\mu&\simeq&-(p_F-p)^2,\ p<p_F,\\
\varepsilon-\mu&\simeq&\,\,\,(p-p_F)^2,\ p>p_F,\nonumber
\end{eqnarray}
As a consequence, the FC state and its corresponding flat bands
emerge beyond the topological FCQPT
\cite{khod:2015,khodel:1994,pr,book}, while the critical
temperature assumes the behavior $T_c\propto \sqrt{x-x_c}$
\cite{abrikosov}. These results are consistent with the
experimental data \cite{bosovic}.

At $T=0$, the onset of FC in homogeneous matter is attributed to
a nontrivial solution $n_0(p)$ of the variational equation
\cite{ks91}
\begin{equation}
\frac{\delta E[n(p)]}{\delta n(p)}-\mu=0 , \ p\in [p_i,p_f].
\label{var}
\end{equation}
Here $E$ is the ground-state energy functional (its variation
generating a single-electron spectrum $\varepsilon$), while
$p_i$ $p_f$ denote the limits of the momentum interval within
which the solution of Eq.~\eqref{var} exists (see
Refs.~\cite{book,pr,ks91} for details).  To be more specific,
Eq.~\eqref{var} describes the flat band pinned to the Fermi
surface, resulting from fermion condensation.

\begin{figure} [! ht]
\begin{center}
\includegraphics [width=0.9\columnwidth]{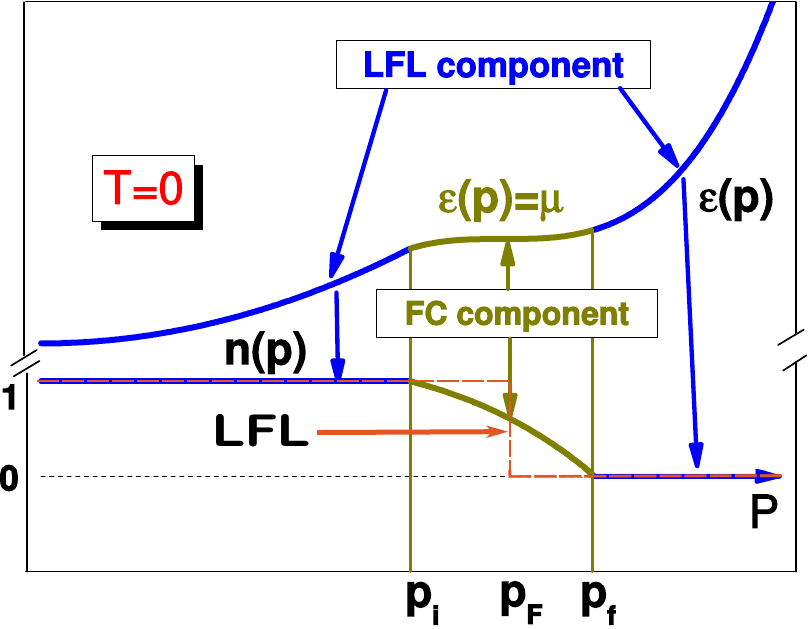}
\end{center}
\caption {(Color online) Schematic representation of
two-component electron liquid at $T=0$ with an FC component. Red
dashed line (marked ``LFL'') shows $n(p)$ for the system {\it
without} a FC, which as ordinary step-function  With the
present, the system is separated into two components: (i) a
normal Fermi liquid with quasiparticle distribution $n(p<p_i)=1$
and $n(p>p_f)=0$. (ii) the FC with $0<n(p_i<p<p_f)<1$ and
single-particle spectrum $\varepsilon (p_i<p<p_f)=\mu$. The
Fermi momentum $p_F$ is situated between $p_i$ and $p_f$.}
\label{fig1}
\end{figure}

To explain the emergent superconductivity at $x\to x_c$, we need
to examine the consequences of the flattening of the
single-particle excitation spectrum $\varepsilon({\bf p})$
(i.e.\ the appearance of a flat band or bands) in
strongly-correlated Fermi systems.  (See
\cite{pr,volovik:2015,book} for recent reviews.) At $T=0$, the
ground state of a system hosting a flat band is degenerate.  The
occupation numbers $n_0({\bf p})$ of single-particle states
belonging to the flat band are continuous functions of momentum
${\bf p}$, in contrast to the standard LFL ``step'' from 0 to 1
at $p=p_F$, shown by the red dashes (color on-line) in
Fig.~\ref{fig1}. Thus at $T=0$ the superconducting order
parameter becomes $\kappa(p)=\sqrt{n(p)(1-n(p))}\neq 0$ in the
region occupied by FC \cite{khodel:1994,pr,qp2}. This property
is in a stark contrast to standard LFL picture, where at $T=0$
and $p=p_F$ the order parameter $\kappa(p)$ is necessarily zero,
as seen in Fig.~\ref{fig1}.  Because of the fundamental
difference between the FC single-particle spectrum and that of
the remainder of the Fermi liquid, a system having FC is, in
fact, a two-component system, separated from the ordinary Fermi
liquid by the driving topological phase transition
\cite{vol3,khodel:1994,volovik:2015}. The range $L$ of momentum
space adjacent to $\mu$ where the fermion condensate resides is
$L\simeq p_f-p_i$, as indicated in Fig.~\ref{fig1}.

\subsection{Superconductivity in systems with FC}

To analyze quantitatively the emergent superconductivity in
question, it is convenient to make use of Gor'kov's formulation
for the Green's functions of a superconductor
\cite{pr,landau9,gorkov}. For this two-dimensional case,
solutions of the Gor'kov equations \cite{pr,landau9,gorkov}
determine the anomalous and regular Green's functions $F^+({\bf
p},\omega)$ and $G({\bf p},\omega)$ of the superconductor:
\begin{eqnarray}
\nonumber F^+({\bf p},\omega)
&=&\frac{-g_0\Xi^*}{(\omega -E({\bf p})+i\,0)(\omega +E({\bf p})-i\,0)},\\
G({\bf p},\omega)&=&\frac{u^2({\bf p})}{\omega -E({\bf
p})+i\,0}+\frac{v^2({\bf p})}{\omega +E({\bf
p})-i\,0}.\label{zui2}
\end{eqnarray}
Here the single-particle spectrum $\varepsilon ({\bf p})$ is
determined by Eq.~\eqref{var} and
\begin{equation}E({\bf p})=\sqrt{\xi^2({\bf
p})+\Delta^2({\bf p})};\,\,\, \frac{\Delta({\bf p})}{E({\bf
p})}=2\kappa({\bf p}), \label{SC4}\end{equation} with $\xi({\bf
p})=\varepsilon({\bf p})-\mu$, where also $\Delta({\bf
p})/E({\bf p})=2\kappa({\bf p})$. The coefficients $v({\bf p})$
and $u({\bf p})$ of the corresponding Bogoliubov transformation
are related to the quasiparticle distribution $n({\bf p})$ and
order parameter $\kappa({\bf p})$ by $u^2({\bf p})=1-n({\bf
p})$, $v^2({\bf p})=n({\bf p})$, and $\kappa({\bf p})=u({\bf
p})v({\bf p})$ \cite{landau9,gorkov}. The gap $\Delta$ and the
function $\Xi$ are given by
\begin{equation}\label{zui3}
\Delta=g_0|\Xi|,\quad i\Xi= \int F^+({\bf
p},\omega)\frac{d\omega d{\bf p} }{(2\pi)^3},
\end{equation}
where $g_0$ is the superconducting coupling constant.
Qualitatively, $F^+({\bf p},\omega)$ has been interpreted as the
wave function of Cooper pairs, and $\Xi$ as the wave function of
the motion of these pairs as a whole. Taking Eqs.~\eqref{SC4}
and \eqref{zui3} into account, we can rewrite Eqs.~\eqref{zui2}
as
\begin{eqnarray}
F^+({\bf p},\omega)&=&-\frac{\kappa({\bf p})}{\omega -E({\bf
p})+i\,0}+\frac{\kappa({\bf p})}{\omega +E({\bf p})-i\,0}, \nonumber \\
G({\bf p},\omega)&=&\frac{u^2({\bf p})}{\omega -E({\bf
p})+i\,0}+\frac{v^2({\bf p})}{\omega +E({\bf
p})-i\,0}.\label{zui8}
\end{eqnarray}
With $g_0\to 0$, one has $\Delta \to 0$, but $\Xi$ and
$\kappa({\bf p})$ remain finite if the spectrum becomes flat,
i.e., $E({\bf p})=0$. In the interval $p_i\leq p\leq p_f$
Eqs.~(\ref{zui8}) then become \cite{book,pr,shagstep}
\begin{eqnarray}
F^+({\bf p},\omega )&=&-\kappa{\bf p})\left[ \frac{1}{\omega
+i\,0}
-\frac{1}{\omega -i\,0}\right], \nonumber  \\
G({\bf p},\omega )&=&\frac{u^2({\bf p})}{\omega
+i\,0}+\frac{v^2({\bf p})}{\omega -i\,0}.\label{zui9}
\end{eqnarray}
The parameters $v({\bf p})$ and $u({\bf p})$ are determined by
the condition that the spectrum should be flat, thus
$\varepsilon({\bf p})=\mu$. It follows from Eqs.~\eqref{SC4} and
\eqref{zui3} that \begin{equation}\label{zui7} i\Xi=\int
F^+({\bf p},\omega )\frac{d\omega d{\bf
p}}{(2\pi)^3}=i\int\kappa({\bf p})\frac{d{\bf
p}}{(2\pi)^2}\simeq n_{FC},
\end{equation}
where $n_{FC}$ is the density of superconducting electrons
forming the FC component (cf.\ Fig.~\ref{fig1}).

We construct the functions $F^+({\bf p},\omega)$ and $G({\bf
p},\omega)$ in the case where the constant $g_0$ is finite but
small, such that the functions $v({\bf p})$ and $\kappa({\bf
p})$ can be found from the FC solutions of Eq.~(\ref{var}). Then
$\Xi$, $\Delta$, and $E({\bf p})$ are given by
Eqs.~(\ref{zui7}), (\ref{zui3}), and (\ref{SC4}), respectively.
Substituting the functions found in this manner into
(\ref{zui8}), we obtain $F^+({\bf p},\omega)$ and $G({\bf
p},\omega)$. We note that Eqs.~\eqref{zui3} and \eqref{zui7}
imply that the gap $\Delta$ is a linear function of both $g_0$
and $n_{FC}$. Since $T_c\sim \Delta$, we conclude that the
critical temperature behaves as $T_c \propto
n_{FC}\propto\rho_s$.  Since we consider the overdoped HTSC case
and FCQPT takes place at $x=x_c$, we also have
\begin{equation}
n_{FC}\propto p_F(p_f-p_i)\propto x_c-x, \label{nfx}
\end{equation}
with $(p_f-p_i)/p_F\ll1$ \cite{qp2,pr}.  Thus we arrive at the
key result
\begin{equation}
n_{FC}=n_s\ll n_{el}.\label{nfc}
\end{equation}
Increasing $g_0$ causes $\Delta$ to become finite, leading to a
finite value of the effective mass $m^*_{FC}$ in the FC state
\cite{pr,qp2}:
\begin{equation}
m^*_{FC}\simeq p_F\frac{p_f-p_i}{2\Delta}.\label{SC7}
\end{equation}

An important fact is warrants emphasis at this point.  It has
been shown in Refs.~\cite{book,pr} that within the FC formalism,
the BCS relations remain valid if we use the spectrum given by
Eq.~\eqref{SC7}. What this means is that the standard BCS
approximation can be used with the a momentum-independent
superconducting coupling constant $g_0$ in the region
$|\varepsilon({\bf p})-\mu|\leq \omega_D$, considering $g_0$ to
be zero outside this region. As usual, $\omega_D$ is a
characteristic energy proportional to the Debye temperature.
With these prescriptions, the superconducting gap depends only
on temperature and is determined by the equation
\cite{book,khodel:1994,pr}
\begin{figure} [! ht]
\begin{center}
\includegraphics [width=1.0\columnwidth]{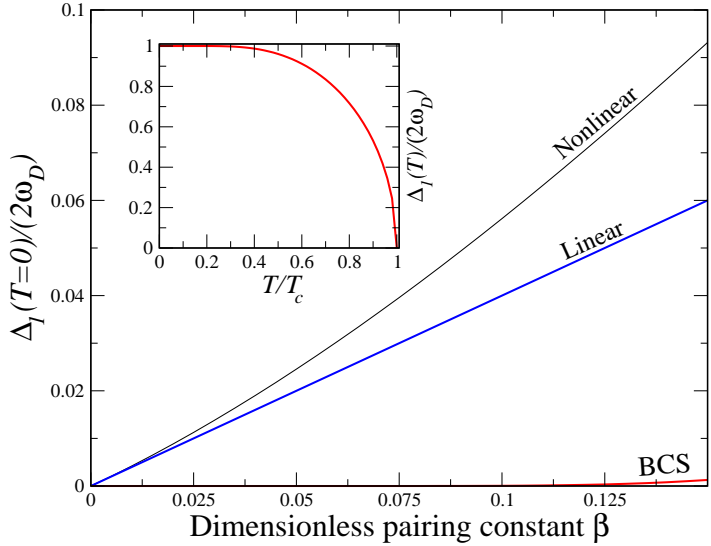}
\end{center}
\caption {(Color online) Solution of Eq.~\eqref{gap2} in the
form $\delta(\beta)$ at $B=0.4$.  The curve labeled
``Nonlinear'' was obtained by direct numerical solution of the
transcendental equation \eqref{gap2}.  The curve labeled
``Linear'' plots the linear dependence $\delta=B \beta$, while
the curve at the bottom shows the BCS dependence resulting from
Eq.~\eqref{gap2} at $B=0$. It is seen that the FC description
allows for much higher critical temperatures $T_c$ proportional
to superconducting gap $\Delta(T=0)$ to be reached than the BCS
estimate.  The inset displays the temperature dependence the
superconducting gap in the FC description at $B=0.4$.}
\label{fig2}
\end{figure}
\begin{eqnarray}
\frac{1}{g_0}=N_{FC}\int_0^{E_0/2}\frac{d\xi}{f(\xi,\Delta)}
\tanh\frac{f(\xi,\Delta)}{2T}+\nonumber \\
+N_L\int_{E_0/2}^{\omega_D}\frac{d\xi}
{f(\xi,\Delta)}\tanh\frac{f(\xi,\Delta)}{2T}, \label{gap1}
\end{eqnarray}
where $f(\xi,\Delta)=\sqrt{\xi^2+\Delta^2(T)}$ and
$E_0=\varepsilon(p_f)-\varepsilon(p_i) \approx 2\Delta (T=0)$ is
a characteristic energy scale. Additionally,
$N_{FC}=(p_f-p_F)p_F/(2\pi \Delta(T=0))$ and $N_L=m^*_L/(2\pi)$
are the densities of states of FC and non-FC electrons,
respectively, $m^*_L$ being the effective mass of electron of
the LFL component, as per Fig.~\ref{fig1}. In the limiting case
$T=0$, we have $\tanh(f/(2T))=1$ as usual and the remaining
integrals can be evaluated exactly. This exercise yields the
following equation relating the value $\Delta (T=0)$ with the
superconducting coupling constant $g_0$:
\begin{equation}\label{gap2}
\frac{\delta}{\beta}=B-\delta \ln \delta,
\end{equation}
where $\beta=g_0m^*_L/(2\pi)$ is a dimensionless coupling
constant, $\delta=\Delta(T=0)/(2\omega_D)$, and
$B=(E_F/\omega_D)((p_f-p_F)/p_F) \ln(1+\sqrt{2})$. The parameter
$B$ is seen to depend on the width of FC interval, so at $B=0$
(i.e., $p_f=p_F$) system is entirely out of the FC and hence in
a pure BCS state.  In this case the solution of \eqref{gap2} has
the standard BCS form $\Delta_{BCS}=\exp(-1/\beta)$. On the
other hand, at small $\beta$ but $B \neq 0$ we obtain a linear
relation between the gap and coupling constant, $\delta=B\beta$.
This not only differs drastically from the BCS result, it also
provides much higher $T_c$ that is directly proportional to
$\Delta(T=0)$. Indeed, for the FC solution one typically has
\cite{book,pr} $T_c \approx \Delta(T=0)/2$ for the linear
regime. Fig.~\ref{fig2} displays results from solution of
Eq.~\eqref{gap2} in the case of $B=0.4$ and small $\beta$, below
0.15.  It is seen that the linear regime of the FC theory
already provides much higher $T_c$ values than BCS; indeed, the
nonlinear treatment giving the complete numerical solution of
Eq.~\eqref{gap2} yields even higher $T_c$. In short, this means
that the FC approach is quite capable of explaining the most
salient feature of high-$T_c$ superconductivity. Referring again
to Fig.~\ref{fig2}, the inset shows a plot of the temperature
dependence of the FC-derived gap corresponding to
Eq.~\eqref{gap1}, in dimensionless units. This dependence is not
specific to the FC approach, as it is qualitatively similar to
that of the BCS case. Moreover, variation of the
``FC-parameter'' $B$ (even setting $B=0$) does not change the
picture qualitatively.

We next analyze the superfluid density $n_s$ for finite $g_0$.
As seen from Eqs.~\eqref{zui7} and \eqref{zui9}, $n_s$ emerges
when $x\simeq x_c$, and occupies the region $p_i\leq p\leq p_f$.
Hence we may write $n_s=n_{FC}\propto x_c-x$, where $n_{FC}$ is
the electron density in FC phase.  It follows that in the FC
phase one has $n_s\ll n_{el}=n_{FC}+n_L$, where $n_{el}$ and
$n_L$ are respectively the total density of electrons and the
electron density outside the FC phase. We should note that the
result $n_s \sim n_{el}$ does is not restricted to BCS theory of
superconductivity, but rather is a much deeper property
belonging to almost any superfluid system, by virtue of the
Leggett theorem \cite{leg}. The practical implication of this
theorem is that at $T=0$ in any system exhibiting superfluidity
(or superconductivity in the case of charged particles), the
number $n_s$ density of particles exhibiting superfluid behavior
should not depart substantially from the number density of all
liquid particles of the system.  In the present case of an
electronic fluid, this means that $n_s \approx n_e$, where $n_e$
is the total electron density.  However, a condition for this
theorem to hold is that the system be $T$-invariant, i.e.,
possess time-reversal symmetry.  Topologically, the FC state is
quite non-trivial topologically \cite{book,
baras,tun,volovik:2015,vol1}, and lacks this symmetry (also
violating $\rm CP$ invariance, i.e., symmetry under combined
charge conjugation and parity reversal \cite{pr,baras}). The
exactly solvable model and general consideration show that the
inequality $n_s\ll n_{el}$ is inherent to FC, as it is seen from
Eqs.~\eqref{zui7} and \eqref{nfc} \cite{qp2,pr,book}. Absent
some even more exotic mechanism, and in the presence of a host
of other favorable experimental observations, it is reasonable
to propose that the main contribution to the superconductivity
in overdoped high-$T_c$ materials as revealed by the recent
experiments \cite{bosovic,zaanen} has its origin in the FC
state. Pairing with such unusual properties can be viewed as a
shadow of fermion condensation -- a situation foretold by an
exactly solvable model long before the experimental observations
were obtained by Bo\^zovi\'c et al. \cite{bosovic} and
demonstrating that both the gap and the order parameter exist
only in the region occupied by fermion condensate \cite{qp2}.
Thus, the experimental observations \cite{bosovic} can be viewed
as a direct experimental manifestation of FC,
 {while another direct
experimental manifestation of FC has been done recently
\cite{mel2016}, where FC phenomenon has been detected in
two-dimensional SiGe/Si/SiGe heterostructures}.

The essential message of the above deliberations is that fermion
condensation entails emergence of a two-component (two-fluid)
system that explains, in a natural way, the deficit $n_s << n_e$
(inconsistent with BCS) that is observed in overdoped high-$T_c$
materials, while allowing for the observed high critical
temperatures.

\subsection{Penetration depth, thermodynamic and transport properties}

We now address the question: does our FC-based superconductor
belongs to the London type. To this end, we recall London's
electrodynamics equations, namely
\begin{equation}
\nabla\times{\bf j}_s=-(n_se^2/m^*){\bf
B}\equiv-(n_{FC}e^2/m^*_{FC}) {\bf B}, \qquad \nabla \times {\bf
B} =4\pi {\bf j}_s,
\end{equation}
where ${\bf j}_s$ is a superconducting current.  These equations
imply a penetration depth
\begin{equation}
\label{lamb} \lambda^2=\frac{m^*_{FC}}{4\pi e^2n_{FC}}.
\end{equation}
Comparing this penetration depth with the coherence length
$\xi_0\sim p_F/(m^*_{FC}\Delta)$, we conclude that $\lambda >>
\xi_0$ since the FC quasiparticle effective mass is huge
\cite{book}. Thus, the superconductors being considered are
indeed of the London type.
\begin{figure} [! ht]
\begin{center}
\includegraphics [width=1.0\columnwidth]{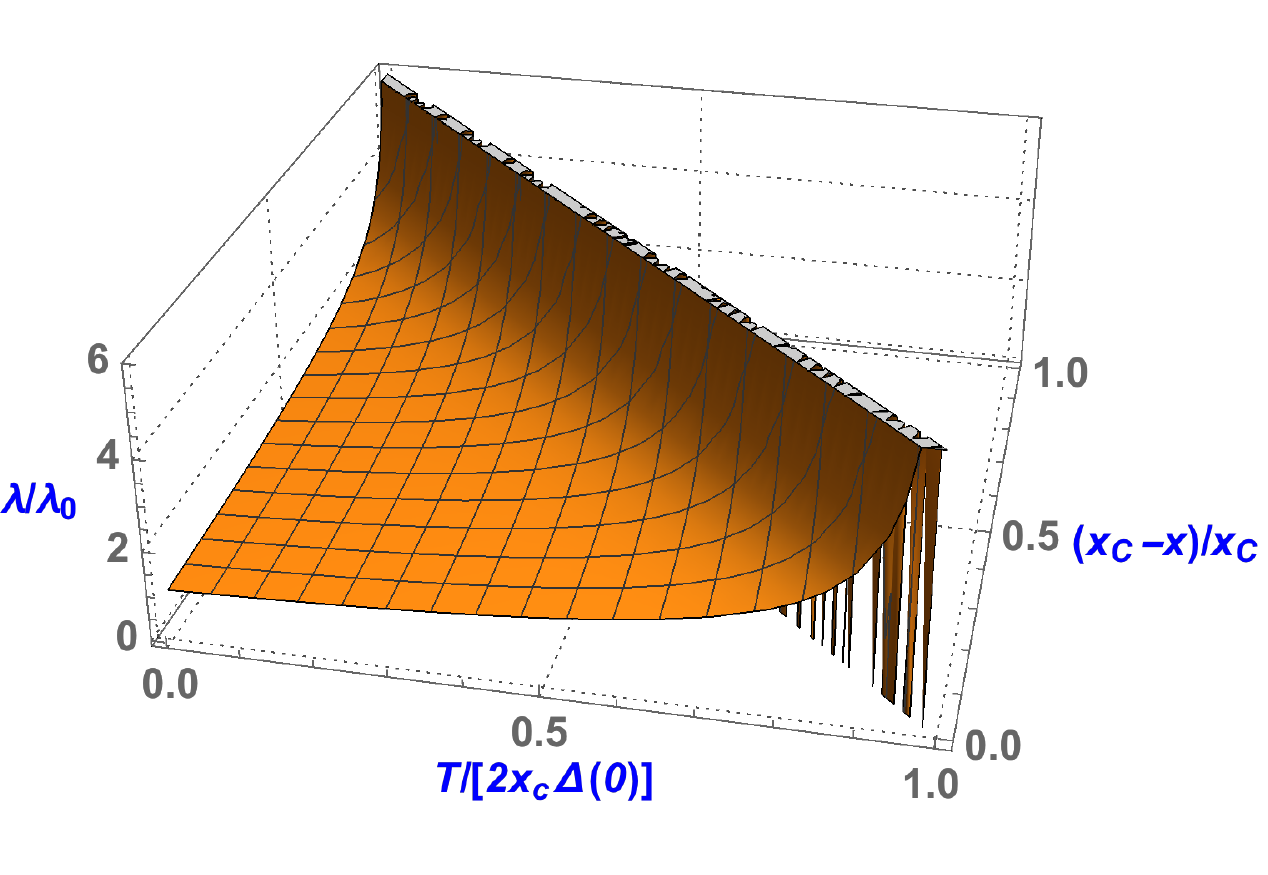}
\end{center}
\caption {(Color online) Dependence of the dimensionless
penetration depth $\lambda/\lambda_0$ of Eq.~\eqref{lamb1} on
temperature and doping.} \label{fig3}
\end{figure}

It turns out that in the FC phase, the penetration depth is a
function not only of temperature, but also of doping degree $x$.
It then follows from Ginzburg-Landau theory that the density of
superconducting electrons $n_s$ grows with $T_c-T$, $n_s\sim
T_c-T$. On the other hand, as has been discussed in
Ref.~\cite{bosovic}, pressure enhances $n_s$, meaning that the
density $x$ of charge carriers is important.  Further, it has
been shown~\cite{book,pr}) that $T_c(x=0)\simeq 2\Delta(T=0)$ in
a superconducting phase involving a FC.  This allows us to use
the relation \eqref{lamb} to plot the penetration depth as a
function of temperature and doping in the form
\begin{equation}\label{lamb1}
\frac{\lambda}{\lambda_0}=\frac{1}{\sqrt{1-y-\tau}},
\end{equation}
where $y=(x_c-x)/x_c$ and $\tau=T/(2\Delta(T=0)x_c)$, with
$\lambda_0$ combining all proportionality coefficients entering
the problem. The dependence \eqref{lamb1} is depicted in
Fig.~\ref{fig3}. Very good qualitative agreement with
experimental data (Fig.~2a from Ref.~\cite{bosovic}) is seen. In
particular, the doping-dependent penetration depth $\lambda$
becomes infinite at the superconducting phase transition
temperature. At zero temperature the divergence of $\lambda$
occurs at $x\simeq x_c$, corresponding to FC-phase emergence,
i.e. at $T=0$ both superconductivity and the FC phase make their
appearance.  At higher temperatures, $\lambda$ becomes divergent
in the region $x<x_c$, i.e.\ deeply inside the FC phase. This
produces ``traces'' of the FC at finite temperatures. In effect,
the present approach, based on the hypothesis of a topological
FC quantum phase transition, describes the most essential and
puzzling features of overdoped HTSC.

The essential features of our dual-component many-fermion model
that equip it for success in explaining the unexpected behavior
of overdoped high-$T_c$ superconductors stem derived from two
distinctive properties of its fermion-condensate component.
First, this FC component necessarily involves only a small
fraction of the traditional Fermi sphere, thus endowing
superfluidity on a number density $n_s \simeq n_{FC}$ of
electrons much smaller than the total electron number density
$n_e$. And second, FC superconductivity entails a relation
between coupling strength and energy gap (hence $T_c$), namely
Eq.~\eqref{gap2}, which is altered drastically from that of BCS
theory, providing for the much higher critical temperatures
observed and predicted.

Another favorable attribute of the two-fluid FC model is
consistency with Uemura's law \cite{uemura}.  Since $T_c\propto
n_s/m^*\equiv n_{FC}/m^*_{FC}$, we may call upon
Eqs.~\eqref{SC7} and \eqref{lamb} to derive
\begin{equation}
\frac{\rho_s}{A}=\lambda^{-2}\simeq\frac {n_s}{m^*}\simeq \frac
{n_{FC}}{m^*_{FC}}\simeq 2\Delta\simeq T_c.\label{SC8}
\end{equation}
Taking into account $n_{\rm FC}\propto x_c-x$, we find that
Eq.~\eqref{SC8} reproduces the main results of our analysis, in
good agreement with the experimental data \cite{bosovic,zaanen}.
The dependence of $T_c$ on $\rho_s$ is seen to be linear, thus
representing the observed scaling law, while $T_c$ is primarily
controlled by $n_s$ \cite{bosovic}. We note that the results for
underdoped HTSC \cite{uemura,uem_n} are similar to those for
overdoped HTSC, which suggests an underdoped/overdoped symmetry
\cite{bosovic}. Consequently, we find good agreement with
Uemura's law in overdoped LSCO as well \cite{bosovic}.

It is worth pointing out that for doping levels $x>x_c$ at which
FCQPT has not yet occurred, the system is in the LFL phase where
the resistivity behaves as $\rho\propto T^2$, indicative of
``more metallic'' character than that exhibited in the FC phase
\cite{bosovic,pagl,khod:2015,jetp2003,arch}.  Superconductivity
as observed appears in the latter phase because the FC
phenomenon strongly facilitates the superconducting state. In
the ``normal'' phase at $T>T_c$, FC gives rise to linear
$T$-dependence of the resistivity, $\rho(T)\propto T$
\cite{qp1,qp2,khod:2015,arch}, in good qualitative agreement
with the experimental data on LSCO and $\rm La_{2-x}Ce_xCuO_4$
\cite{bosovic,pagl}. We note that in the transition region
$x\simeq x_c$, the behavior $\rho(T)\propto T^{\alpha}$ is
observed with $\alpha\sim 1.0-2.0$ \cite{pagl,khod:2015,arch}.

\section{Conclusions}\label{SUM}

Combining analytical considerations with arguments based
entirely on experimental grounds, we have shown that data
collected on very different strongly correlated many-fermion
systems demonstrates a remarkable commonality among them, as
expressed in universal scaling behavior of their thermodynamic,
transport, and relaxation properties, independently of the great
diversity in their individual microstructure and microdynamics.
The systems considered range from heavy-fermion metals, to
quantum liquids including $^3$He films, to insulating compounds
possessing one-, two-, and three-dimensional quantum spin-liquid
states, to quasicrystals and beyond.  The universal behavior
exhibited by this class of systems, generically known as
heavy-fermion (HF) systems or compounds, being analogous to that
commonality expressed in gaseous, liquid, and solid states of
matter, leads us to consider such HF systems as manifestations
of a new state of matter arising from the presence of flat bands
in their excitation spectra.  Such flat bands arise from the
formation of a fermion condensate (FC) due to a specific quantum
phase transition, as it is foretold in 1990 \cite{ks91}.

In order to facilitate experimental observation of the FC state
in trapped, ultracold atomic gases, we have formulated and
solved a simple yet realistic model that predicts the appearance
of fermion-condensation precursors in a two-dimensional ensemble
of ultracold fermionic atoms interacting with coherent resonant
light. We have shown that thermodynamic characteristics of the
system exhibit experimentally observable signatures of
FC-precursor realization. Such features can be regarded as
fermion-condensation fingerprints in the system under
consideration.  {We note that the direct experimental
manifestation of fermion condensation has been done recently
\cite{mel2016}, where FC phenomenon has been detected in
two-dimensional SiGe/Si/SiGe heterostructures.}

We have concluded our study of exemplifications of the new state
of matter reached by fermion condensation with an exploration of
high-$T_c$ superconductors as potential hosts of fermion
condensates.  In fact, we have shown that the underlying
physical mechanism responsible for the unusual properties of the
overdoped compound $\rm La_{2-x}Sr_xCuO_4$ (LSCO) observed
recently \cite{bosovic,zaanen} may very well involve a
topological quantum phase transition that induces fermion
condensation. Since the topological FC state violates
time-reversal symmetry, the Leggett theorem no longer applies.
Instead, we have demonstrated explicitly that the superfluid
number density $n_s$ turns out to be small compared to the total
number density of electrons. We have also shown that the
critical temperature $T_c$ is a linear function of $n_s$, while
$n_s(T)\propto T_c-T$. Pairing with such unusual properties is
as a shadow of fermion condensation -- a situation foretold by
an exactly solvable model \cite{qp2} long before the
experimental observations were obtained by Bo\^zovi\'c et al.
\cite{bosovic} and demonstrating that both the gap and the order
parameter exist only in the region occupied by fermion
condensate. Thus, the experimental observations \cite{bosovic}
can be viewed as a direct experimental manifestation of FC.
Additionally, we have demonstrated that at $T>T_c$ the
resistivity $\rho(T)$ varies linearly with temperature, while
for $x>x_c$ it exhibits metallic behavior, $\rho(T)\propto T^2$.
Thus, pursuit of a superconductivity formalism adapted to the
presence of a fermion condensate captures all the essential
physics of overdoped LSCO and successfully explains its most
puzzling experimental features, thereby allowing us to close the
colossal gap existing between the experiments and
Bardeen-Cooper-Schrieffer-like theories. Indeed, these findings
are applicable not only to LSCO but also for any overdoped
high-temperature superconductor.

\begin{acknowledgements}
We are grateful to V.A. Khodel for valuable discussions. This
work was partly supported by U.S. DOE, Division of Chemical
Sciences, Office of Basic Energy Sciences, Office of Energy
Research. JWC is indebted to the University of Madeira and its
Centro de Ci\'encias Matem\'aticas for gracious hospitality
during his sabbatical residency.
\end{acknowledgements}


\end{document}